\documentstyle[12pt,epsf,graphicx,epsfig]{article}
\newcommand{\be}{\begin{eqnarray}}

\newcommand{\ee}{\end{eqnarray}}

\makeindex

\begin{document}

\title{What is the Evidence for the Color Glass Condensate?
\footnote{Invited talk at the NATO Advanced Study Institute:
Structure and Dynamics of Elementary Matter, Kemer, Turkey,
Sep. 22 - Oct. 2, 2003} }
\author{Larry McLerran\\
     { \small\it Physics Department 
PO Box 5000} \\ {\small\it Brookhaven National Laboratory} \\  
{\small\it Upton, NY 11973 USA }\\
 }

\maketitle

\abstract{ I introduce the concept of the Color Glass Condensate.  I 
review data from HERA and RHIC which suggest that such a universal form of
matter has been found.}

\section{What is the Color Glass Condensate?}

The ideas for the Color Glass Condensate originate in the result for the
HERA data on the gluon distribution function shown in 
\begin{figure}[htb]
    \centering
       \mbox{{\epsfig{figure=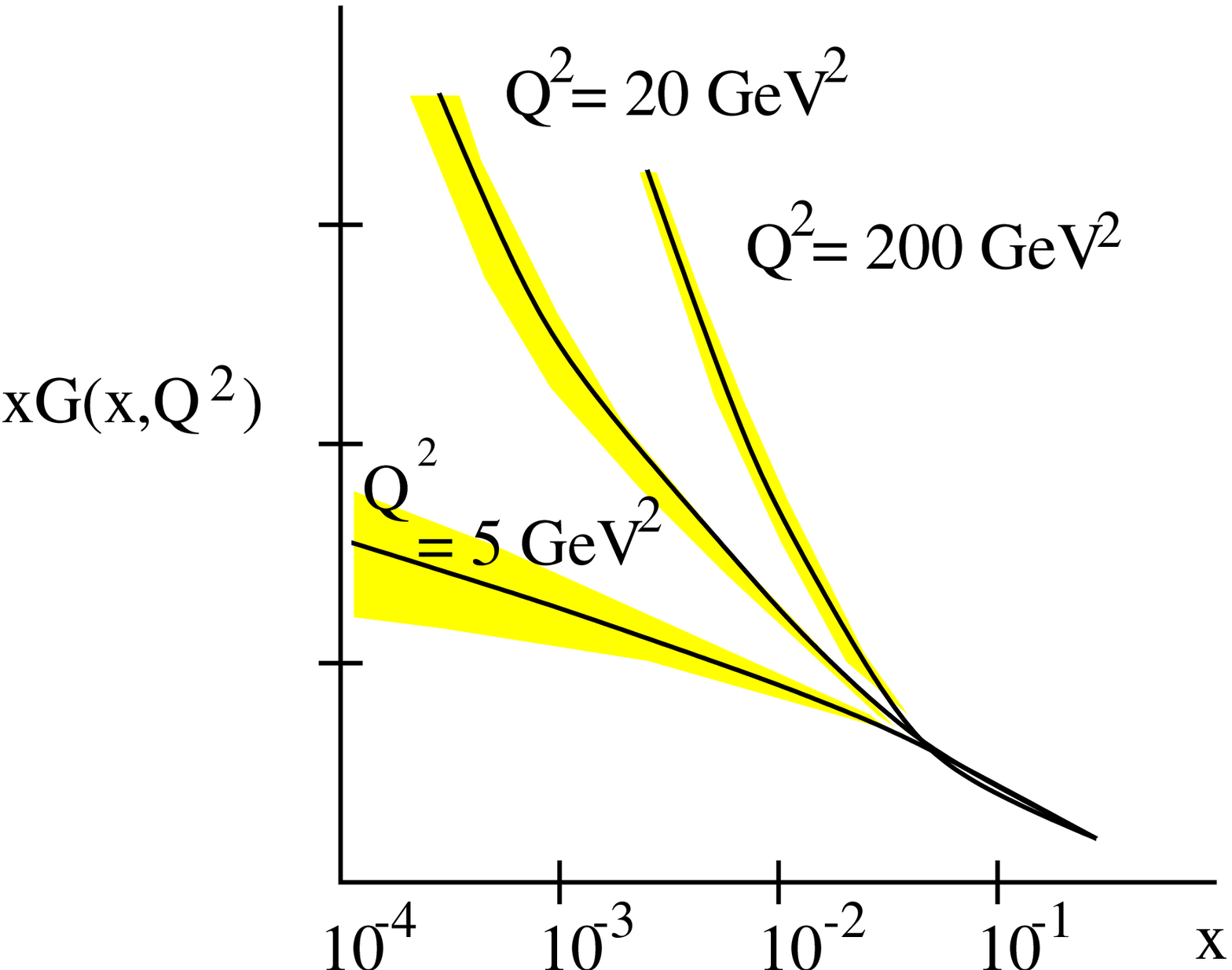,
        width=0.50\textwidth}}\quad
             {\epsfig{figure=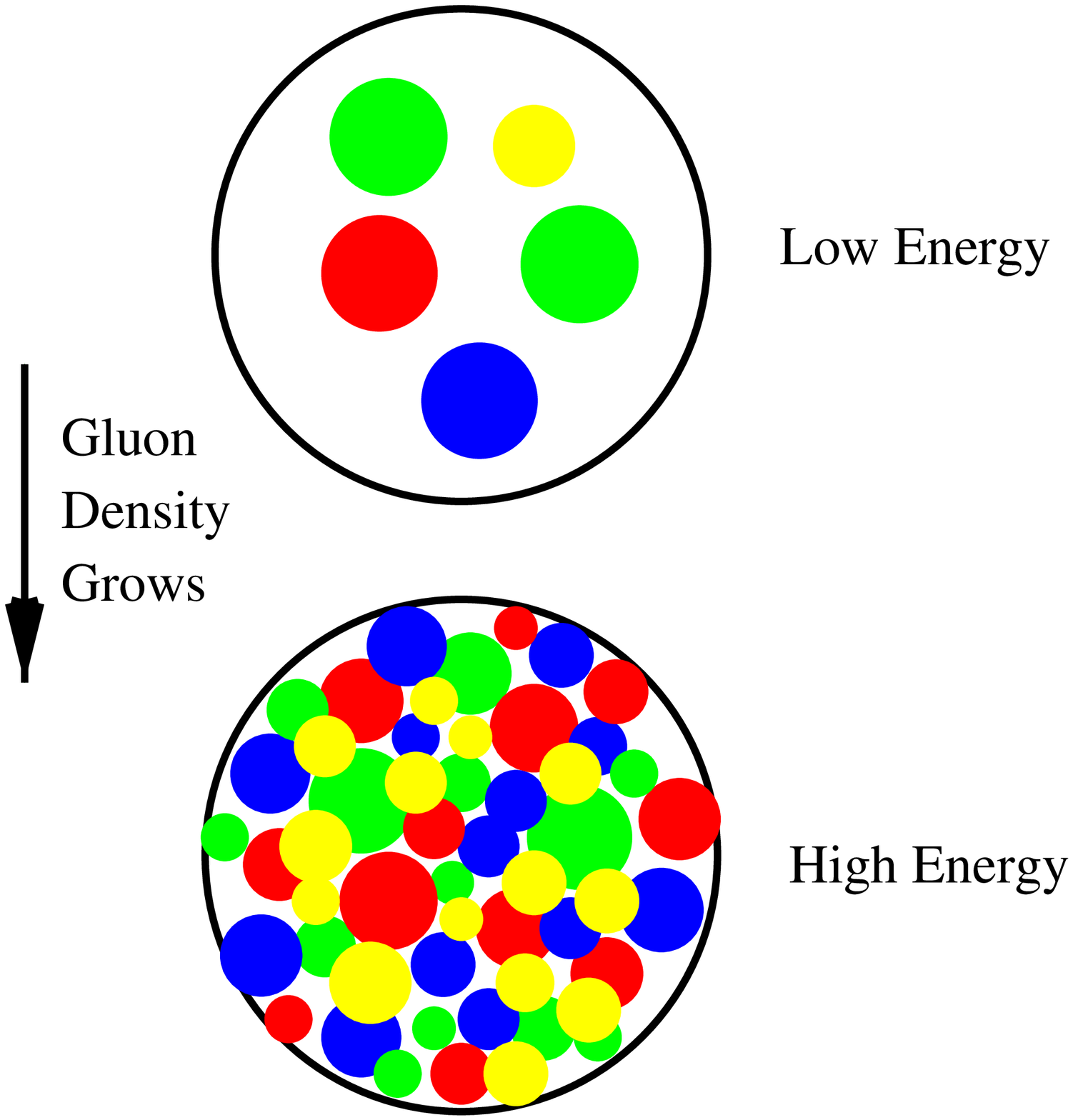,
        width=0.50\textwidth}}}
        \caption{(a)The HERA data for the gluon distribution function
as a function of x for various values of $Q^2$. (b) 
A physical picture of the low x gluon density inside a hadron
as a function of energy }
        \label{heradata}
\end{figure}
Fig. \ref{heradata}(a)
\cite{hera}  The gluon density is rising rapidly as a function of
decreasing x.  This was expected in a variety of 
theoretical works,\cite{glr}-\cite{bfkl} and has the implication that the 
real physical transverse density of gluons must 
increase.\cite{glr}-\cite{mq},\cite{mv}.  This follows because total cross 
sections rise slowly at high energies but the number of gluons is rising 
rapidly.  This is shown in Fig. \ref{heradata}(b). This led to
the conjecture that the density of gluons should become limited,
that is, there is gluon saturation. \cite{glr}-\cite{mq}, \cite{mv}

The low x gluons therefore are closely packed together. The 
strong interaction 
strength must become weak, $\alpha_S \ll 1$.  Weakly coupled systems 
should be possible to understand from first principles in $QCD$.\cite{mv}  

This weakly coupled system is called a Color Glass Condensate for reasons 
we now enumerate:\cite{ilm}
\begin{itemize}
\item {\bf Color}  The gluons which make up this matter are colored.
\item{\bf Glass} The gluons at small x are generated from gluons at larger
values of x.  In the infinite momentum frame, these larger momentum gluons
travel very fast and their natural time scales are Lorentz time dilated.  
This time dilated scale is transferred to the low x degrees of freedom
which therefore evolve very slowly compared to natural time scales.  This
is the property of a glass.
\item{\bf Condensate} The phase space density 
\be
\rho = {1 \over {\pi R^2}}{{dN} \over {dyd^2p_T}}
\ee
is generated by a trade off between a negative mass-squared term linear in 
the density which generates the instability, $-\rho$ and an interaction
term $\alpha_S \rho^2$ which stabilizes the system at a phase space density 
$\rho \sim 1/\alpha_S$.  
Because $\alpha_S << 1$,
this means that the quantum mechanical states of the 
system associated with the condensate are multiply occupied.  They are 
highly coherent, and share some properties of Bose condensates.
The gluon occupation factor is very high, of order $1/\alpha_S$, but it is
only slowly (logarithmically) increasing when further increasing the
energy, or decreasing the transverse momentum. This provides saturation
and cures the infrared problem of the traditional BFKL approach.\cite{im2001}

\end{itemize}

Implicit in this definition is a concept of fast gluons which act as sources 
for the colored fields at small x.  These degrees of freedom are treated 
differently than  the fast gluons which are taken to be sources.  
The slow ones are fields.  There is
an arbitrary $X_0$ which separates these degrees of freedom.  This 
arbitrariness is cured by a renormalization 
group equation which requires that physics be independent of
$X_0$.  In fact this equation determines
much of the structure of the resulting theory as its solution flows to
a universal fixed point.\cite{ilm}-\cite{jklw}

\begin{figure}[ht]
    \begin{center}
        \includegraphics[width=0.50\textwidth]{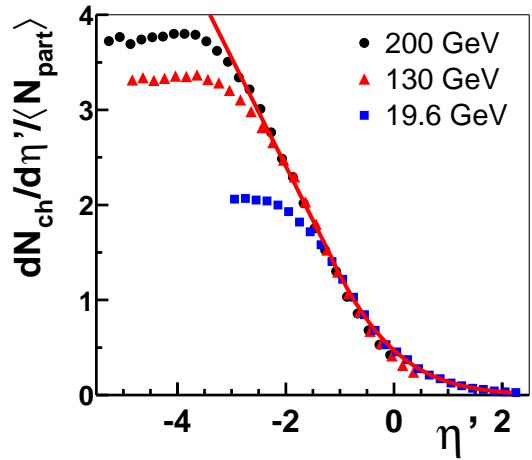}
        \caption{Limiting fragmentation and the RHIC data. }
\label{limfrag}
    \end{center}
\end{figure}

There is evidence which supports this picture.  One piece
is the observation of limiting fragmentation.
This phenomena is that if particles collide at some fixed 
center of mass energy
and the distribution of particles are measured as a function of their 
longitudinal momentum
from the longitudinal momentum of one of the colliding particles, 
then these distributions do not change as one goes to 
higher energy, except for  the new degrees of freedom that appear.
This is true
near zero longitudinal momentum in the center of mass frame because
new degrees of freedom appear as the center of mass energy is increased.  
In the analogy with the CGC,
the degrees of freedom, save the new ones added in at low longitudinal 
momentum, are the sources.  The fields correspond to the new degrees of
freedom.   The sources are fixed in accord with limiting fragmentation.
One generates an effective theory for the low longitudinal
momentum degrees of freedom as fixed sources above some cutoff, and
the fields generated by these sources below the cutoff.  A recent
measurement of limiting fragmentation comes from the Phobos
experiment at RHIC shown in Fig. \ref{limfrag} \cite{phoboslfrag}

Of course the perfect scaling of the limiting fragmentation curves is
only an approximation.  As shown by Jalilian-Marian, 
the limiting fragmentation
curves are given by the total quark, antiquark and gluon distribution 
functions of the fast particle measured at a momentum scale $Q_{sat}^2$
appropriate for  the particle that it collides with.\cite{jamal}  
The saturation momentum $Q_{sat}$
will play a crucial role in our later discussion.  It is a momentum scale
which is determined by the density of gluons in the CGC
\be
	{1 \over {\pi R^2}} {{dN} \over {dy}} \sim {1 \over \alpha_S} Q_{sat}^2
\ee
The saturation momenta 
turns out to depend on the total beam energy because
the longitudinal momentum scale of the target particle at fixed $x$
of the projectile will depend upon the beam energy.  It is nevertheless
remarkable how small these violations appear to be.

The CGC may be defined mathematically by a path integral:
\be
	Z = \int_{X_0} [dA][dj] exp\left(iS[A,j] - \chi[j] \right)
\ee
What this means is that there is an effective theory defined below some cutoff 
in $x$ at $X_0$, and that this effective theory is a gluon field in the 
presence of an external source $j$. This source arises from the quarks
and gluons with $x \ge X_0$, and is a variable of integration.
The fluctuations in $j$
are controlled by the weight function $\chi[j]$.  It is $\chi[j]$
which satisfies renormalization group equations which make the theory
independent of $X_0$.\cite{jkmw}-\cite{jimwlk},\cite{ilm}.  The equation for
$\chi$ is called the JIMWLK equation.  This equation reduces in 
appropriate limits to the BFKL and DGLAP evolution
equations.\cite{bfkl}, \cite{dglap} 
The theory above is mathematically very similar to that of spin glasses.

There are a variety of kinematic regions where one can find
solutions of the renormalization group equations which have
different properties.  There is a region where the gluon density
is very high, and the physics is controlled by the CGC.  This is when
typical momenta are less than a saturation momenta which depends on $x$,
\be
	Q^2 \le Q_{sat}^2(x)
\ee
The dependence of $x$ has been evaluated by several authors,
\cite{glr},\cite{lt}-\cite{mt}, 
and in the energy range appropriate for current experiments
has been determined by Triantafyllopoulos to be 
\be
	Q_{sat}^2 \sim (x_0/x)^\lambda~GeV^2
\ee
where with about $15\%$ uncertainty $\lambda = 0.3$.  The value of $x_0$
is not determined from the renormalization group equations and must be 
found from experiment.

\begin{figure}[ht]
    \begin{center}
        \includegraphics[width=0.50\textwidth]{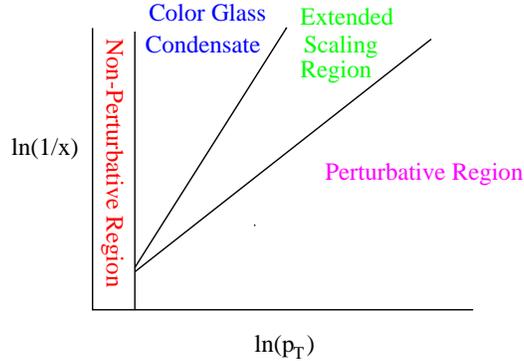}
        \caption{The various regions where there are a
CGC, extended scaling, and low density of glue. }
\label{rhiccgc}
    \end{center}
\end{figure}

There is also a region of very high $Q^2$ at fixed x, where the density of
gluons is small and perturbative QCD is reliable.  It turns out there is
a third region intermediate between high density and low where there are
universal solutions to the renormalization group equations and scaling
in terms of $Q_{sat}^2$.\cite{iim}  In this region and in the region
of the CGC, distribution functions are universal functions of only 
$Q^2/Q_{sat}^2(x)$.  The extended scaling region is when
\be
	Q_{sat}^2 \le Q^2 \le Q_{sat}^4/\Lambda_{QCD}^2
\ee
These various regions are shown in Fig. \ref{rhiccgc} in the plane of
$ln(1/x)$ and $ln(Q^2)$

\section{Why is the CGC Important?}

{\noindent \bf The Color Glass Condensate is  a new universal type of matter:}
\begin{itemize}
\item{\bf Matter:}  The separation between the gluons in the CGC is small
compared to the size of the system.  Due to Lorentz time dilation, the 
lifetime of this matter is long compared to natural time scales.

\item{\bf New:}  This matter can only be probed at high energy, and it may be
produced in ultrarelativistic nuclear collisions.

\item{\bf Universal:} The CGC is universal independent of the type of
hadron which generated it.  Universality of this matter implies it is of
fundamental interest.

\end{itemize}

{\noindent \bf The Color Glass Condensate is a theory of:}

\begin{itemize}
\item{The origin of glue and sea quarks in hadrons.}
\item{The origin and nature of cross sections and particle production.}
\item{The distribution of valence quantum numbers at small x.}
\item{The initial conditions for the matter which evolves into the
Quark Gluon Plasma at RHIC.}
\end{itemize}

\section{What does a high energy hadron look like?}

\begin{figure}[ht]
    \begin{center}
        \includegraphics[width=0.25\textwidth]{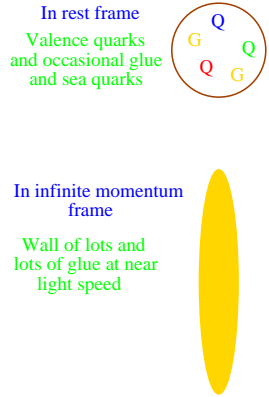}
        \caption{A hadron is shown at rest and in the infinite momentum frame.
 }
\label{frames}
    \end{center}
\end{figure}

In Fig. \ref{frames}, a hadron is shown in its rest frame
and as viewed from the infinite momentum frame.  The picture does not look 
Lorentz invariant.  This is because the hadron is made of many 
different Fock space components, some containing valence quarks and a
few gluons and sea quarks and some containing valence quarks plus many
sea quarks and gluons.  At low energies, one is sensitive to matrix elements
which involve the valence quarks and a few gluons and sea quarks.  For high 
energy collisions, typical matrix elements involve the valence quarks
plus many sea quarks and gluons.  At high energies, the hadron appears as 
if it is a gluon wall.

Of course the Color Glass Condensate description is Lorentz invariant.
In fact, there is a very subtle duality of description.  If one views the 
hadron in the infinite momentum frame, one is scattering from high density
of gluonic matter. In the rest frame of the hadron, where the probe has very
high energy, the Color Glass Condensate appears through coherent multiple
scattering of the probe  valence quarks.  One can prove these 
descriptions are mathematically equivalent.\cite{mue}-\cite{mk} 

\begin{figure}[ht]
    \begin{center}
        \includegraphics[width=0.50\textwidth]{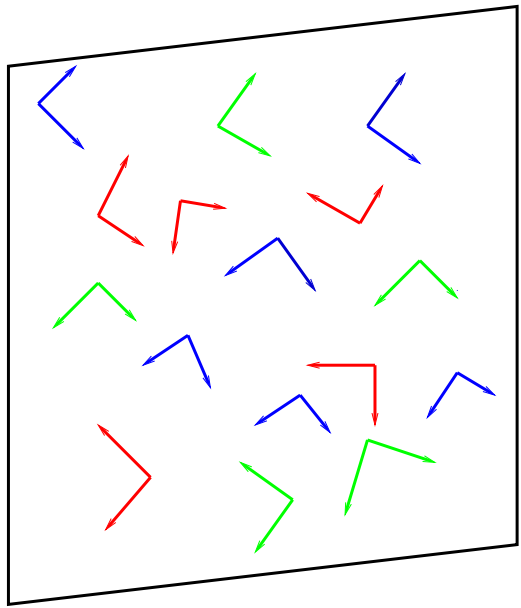}
        \caption{The distribution of colored fields within the CGC.
 }
\label{colorglass}
    \end{center}
\end{figure}

The form of the fields associated with the CGC can also be easily seen.
In the infinite momentum frame, the entire hadron is Lorentz contracted
into a small region of $x^- = t-z$.  The glassy nature of the fields
makes them independent of the light cone time $x^+ = t+z$.  Therefore the
only large component of $F^{\mu \nu}$ is $F^{i +} \sim \partial/\partial x^-
A^i$.  A little algebra shows that 
\be
	\vec{E} \perp \vec{B} \perp \vec{z}
\ee
In Fig. \ref{colorglass}, these fields are shown.  They  correspond to the 
Lienard-Wiechart potential for a boosted electron, except that they are
colored and they have a random polarization and color.

The density of gluons per unit area defines a momentum scale, the
saturation momenta,
\be
	{1 \over {\pi R^2}} {{dN} \over {dy}}
\sim {1 \over \alpha_S} ~Q_{sat}^2(x)
\ee
We insert the extra factor of $1/\alpha_S$ because the fields are
classical, and therefore their density should scale in this way.  This form is
guaranteed because the system is almost scale invariant.  Small violations
of scaling arise because $\alpha_S$ is measured at $Q_{sat}$. 

\section{Experimental Evidence in Support of CGC}

In this section, I discuss the accumulated evidence from HERA and RHIC,
and elsewhere, in support of the hypothesis of a Color Glass Condensate.

\subsection{Geometrical Scaling}

\begin{figure}[ht]
    \begin{center}
        \includegraphics[width=0.50\textwidth]{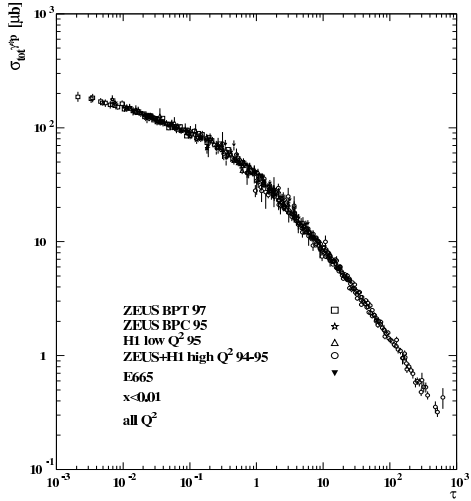}
        \caption{The cross section $\sigma^{\gamma^*p}$ as a function
of the scaling variable $\tau = Q^2/Q_{sat}^2$.
 }
\label{gb}
    \end{center}
\end{figure}

Geometrical scaling is the observation\cite{biel}-\cite{gbks}
that the deep inelastic cross section for virtual photon scattering
as a function of $Q^2$ and $x$ is really only a function of
\be
	\sigma^{\gamma^* p} \sim F(Q^2/Q_{sat}^2)
\ee
where the saturation momentum is taken to be
\be
	Q_{sat}^2 \sim (x_0/x)^\lambda ~ 1 GeV^2
\ee
and $\lambda \sim 0.3$ and $x_0 \sim 10^{-4}$.  This scaling works 
for $x \le 10^{-2}$ and for the available data in $Q^2$.  The data 
is shown in Fig. \ref{gb}

It is straightforward to understand why this scaling works for the small
$Q^2 \le Q_{sat}^2$.  This is the region of the CGC, and there is only
one dimensionful scale which characterizes the system: the saturation 
momentum.\cite{lt}  The surprise is that there is an extended
scaling window for 
$Q_{sat}^2 \le Q^2 \le Q_{sat}^4/\Lambda^2_{QCD}$.\cite{iim} 
This can be proven analytically.  As well, one now has reliable
computation of the dependence on $x$ of the saturation momentum, that is,
one knows the exponent $\lambda$ to about $15\%$ accuracy, and it agrees
with what is seen from the geometrical scaling curve.\cite{mt}  
What is not determined
from the theory of the CGC is the scale $x_0$, and this must be found by
experiment.  This comes from the boundary conditions for the
renormalization group equations.
\begin{figure}[ht]
    \begin{center}
        \includegraphics[width=0.80\textwidth]{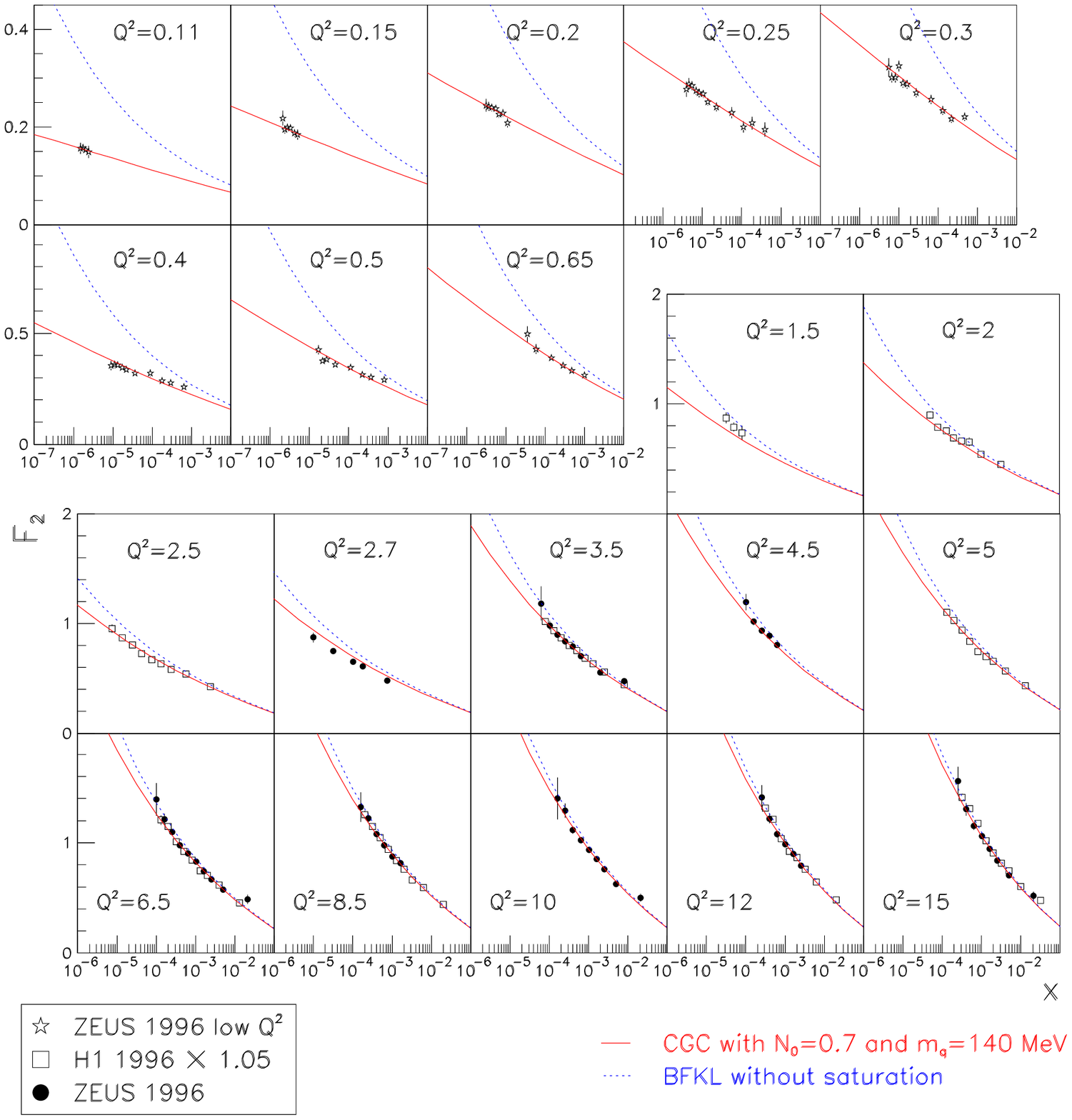}
        \caption{The CGC description of $F_2$.
 }
\label{f2a}
    \end{center}
\end{figure}
\begin{figure}[ht]
    \begin{center}
        \includegraphics[width=0.80\textwidth]{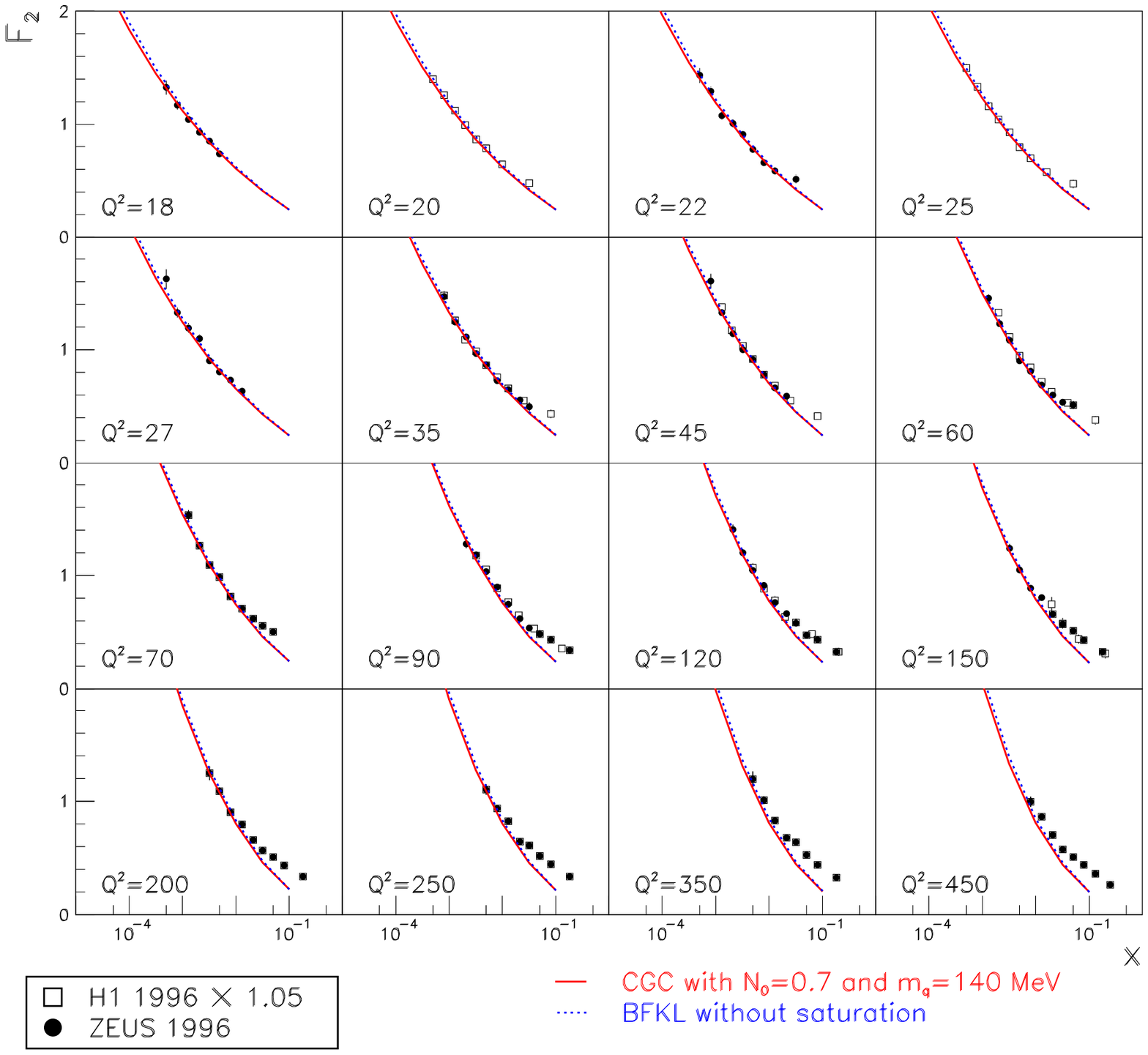}
        \caption{More of the CGC description of $F_2$.
 }
\label{f2b}
    \end{center}
\end{figure}
\subsection{The Structure Function $F_2$}

Using the dipole description of the virtual photon wavefunction,
the structure function $F_2$ can be related to the gluon distribution
function which arises from the CGC.  One can compute and compare to data.
There are 3 unknown parameters in this description:
the hadron size, the scale $x_0$ and the quark mass. 
In addition, the 
parameter $\lambda$ which controls the energy dependence of the
saturation momentum is determined by experiment to better accuracy
than it is currently known theoretically.\cite{lt}-\cite{iimu}
The results for the description of the data are remarkably good
for $x \le 10^{-2}$ and $Q^2 \le 45 ~GeV^2$, as shown below 
in Figs. \ref{f2a}-\ref{f2b}\cite{iimu}

One should note that this description includes both the high and low
$Q^2$ data.  Descriptions based on DGLAP evolution can describe the large
$Q^2$ points.  The CGC description is very economical in the number
of parameters which are used.

\subsection{Diffraction and Quasi-Elastic Processes}

The CGC provides a description of the underlying structure of gluonic
matter inside a hadron.  As such, it should be sensitive to probes of 
the transverse extent of this matter, which can be
experimentally studied in  diffraction and related 
quasi-elastic particle production.\cite{strikman}

\begin{figure}[ht]
    \begin{center}
        \includegraphics[width=0.50\textwidth]{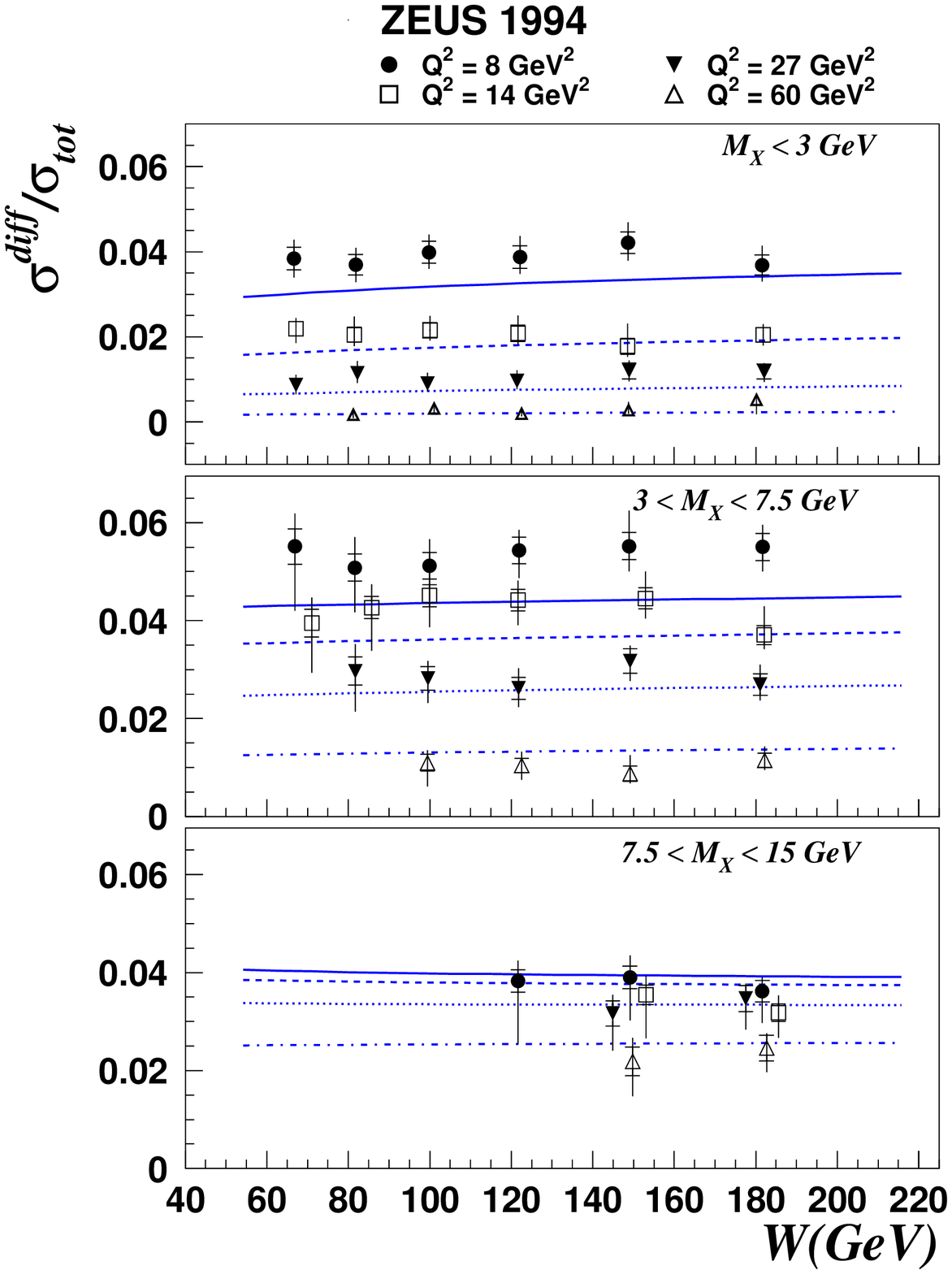}
        \caption{The ratio of diffractive to total cross
sections for deep inelastic scattering.
 }
\label{diff}
    \end{center}
\end{figure}

For diffraction by a virtual photon, a good theoretical first 
principles computation is only 
available for small mass states which are produced by the virtual 
photon.\cite{gbw}-\cite{kowt}.  In Fig. \ref{diff},
a computation of the ratio of the diffractive to total deep inelastic cross
section for various $Q^2$ and produced masses is shown.  The agreement is good
for small masses, and even reasonable for large masses.

\begin{figure}[ht]
    \begin{center}
        \includegraphics[width=0.50\textwidth]{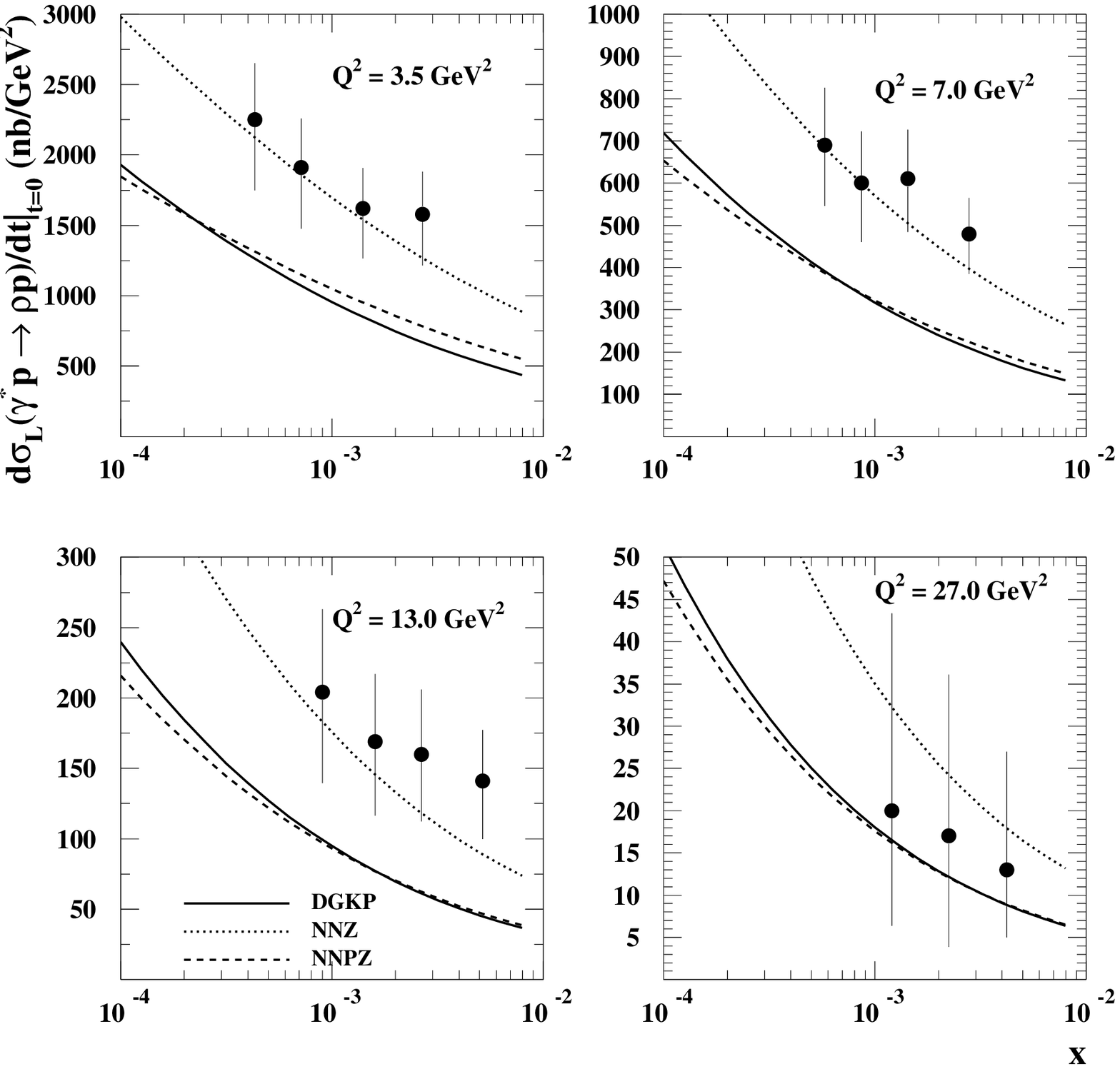}
        \caption{The CGC description of quasi-elastic $\rho$ meson
production.
 }
\label{rho}
    \end{center}
\end{figure}

\begin{figure}[ht]
    \begin{center}
        \includegraphics[width=0.50\textwidth]{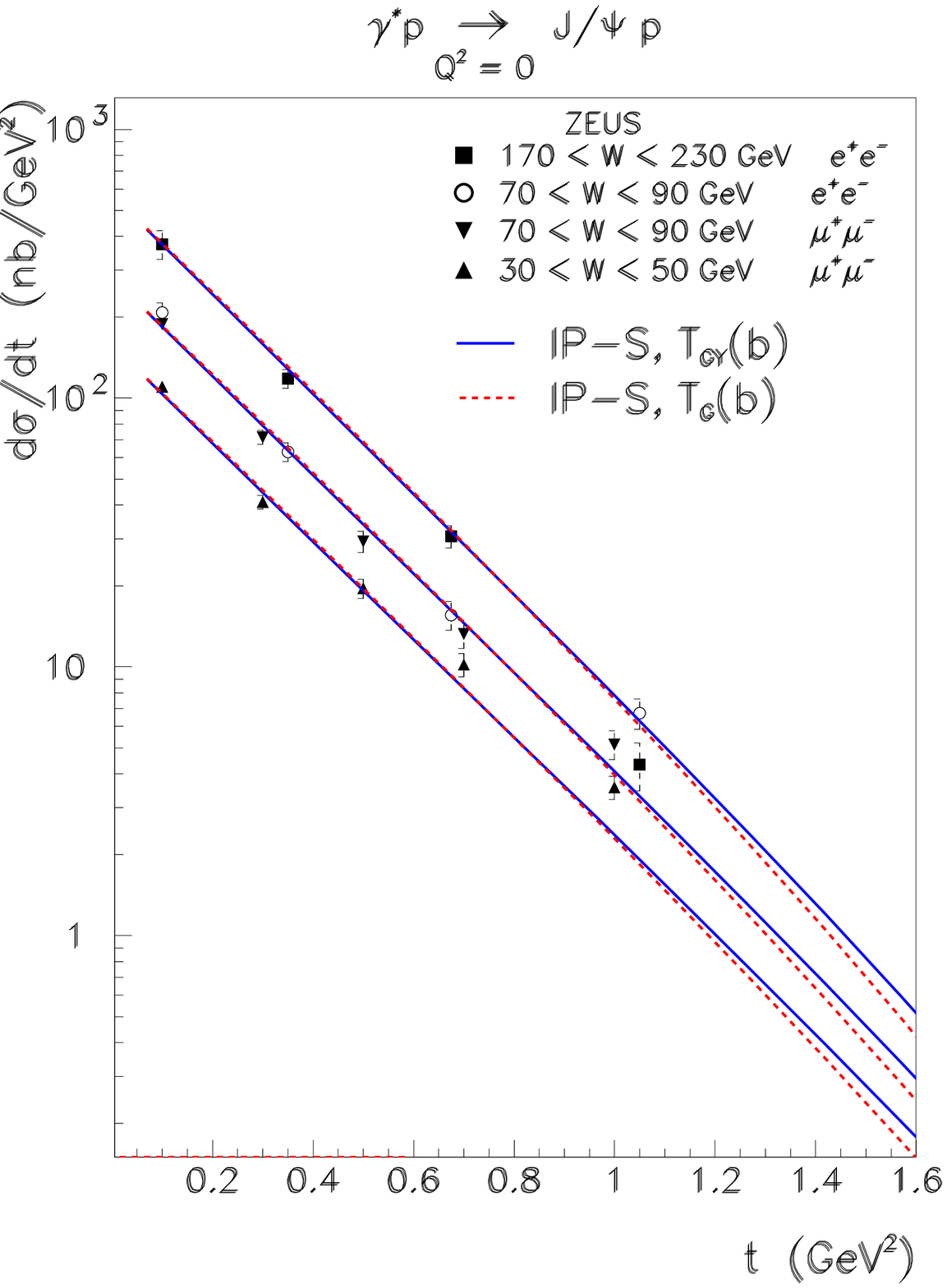}
        \caption{Quasi-elastic $J/\Psi$ production.
 }
\label{jpsi}
    \end{center}
\end{figure}

There are additional computations of quasi-elastic $\rho$ meson 
production and $J/\Psi$ production\cite{mms}-\cite{kowt}.  These are shown in
Figs. \ref{rho}-\ref{jpsi}.  They agree well up to an overall normalization
uncertainty for the $\rho$ meson associated with knowledge of
the $\rho$ meson wavefunction.  The various curves correspond in
the figures correspond to different models for these wavefunctions.


\subsection{Qualitative Understanding of Total Hadronic Cross Sections}

\begin{figure}[ht]
    \begin{center}
        \includegraphics[width=0.50\textwidth]{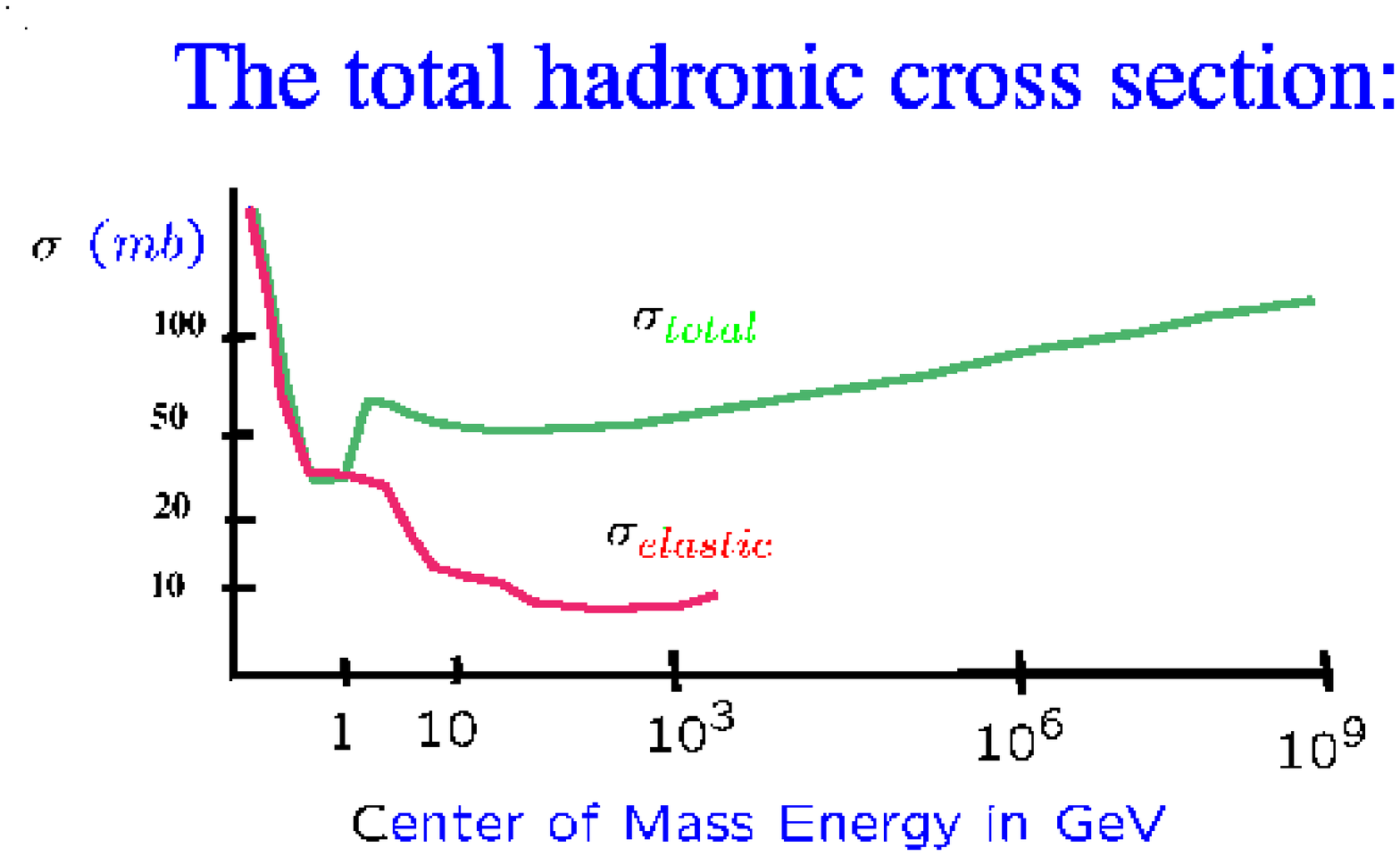}
        \caption{The total $pp$ cross section.
 }
\label{sigtot}
    \end{center}
\end{figure}

The elastic and total cross section of $pp$ scattering 
as a function of energy is shown in
Fig. \ref{sigtot}.
The cross section is slowly varying as a function of energy and is
believed to grow as $ln^2(E)$.  To understand this behaviour, imagine that
we take some probe and try to penetrate the hadron.  The cross section
is defined by the impact parameter at which the hadron become opaque.
We expect that at fixed $x$, the impact parameter distribution of matter
inside a hadron fall off at large $b$ like $e^{-2m_\pi b}$.  At fixed
impact parameter, we expect that the number of gluons grows as 
$(x_0/x)^\lambda$.  Setting
\be
	(x_0/x_{min})^\lambda ~ exp\left(-2m_\pi b \right) \sim 1
\ee
gives
\be
	\sigma \sim b^2 \sim ln^2(1/x_{min}) \sim ln^2(E)
\ee
Here $x_{min}$ is the minimal value of $x$ accessible for some energy
$E$ and goes as $x_{min} \sim \Lambda_{QCD}/E$.

This simple physically motivated picture give the expected 
$ln^2(E)$ growth of the total cross section.  It has proven difficult to make
these simple arguments rigorous.\cite{iimf}-\cite{kw}.

\section{Heavy Ion Collisions}

\begin{figure}[ht]
    \begin{center}
        \includegraphics[width=0.50\textwidth]{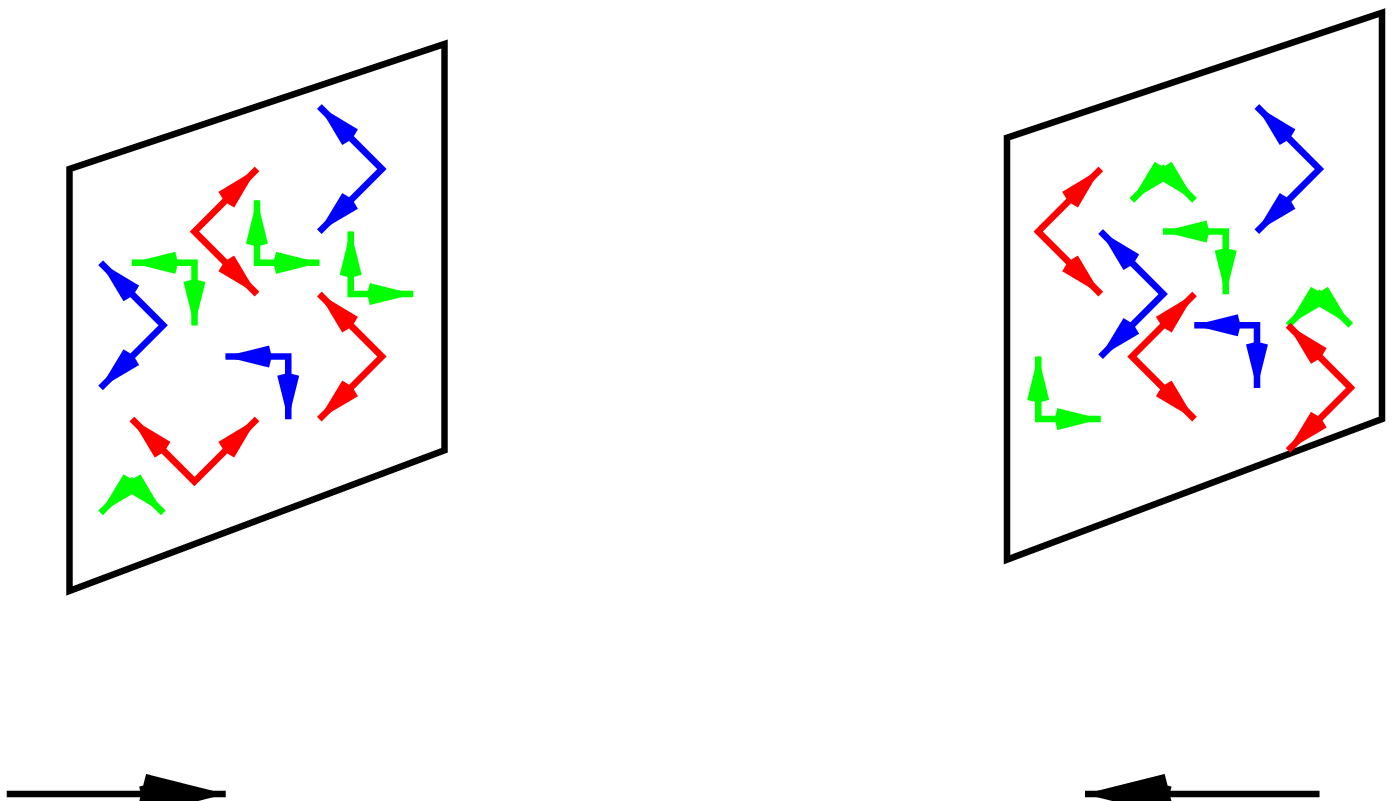}
        \caption{A collision of two ultra-relativistic nuclei.
 }
\label{glasscoll}
    \end{center}
\end{figure}

\begin{figure}[ht]
    \begin{center}
        \includegraphics[width=0.50\textwidth]{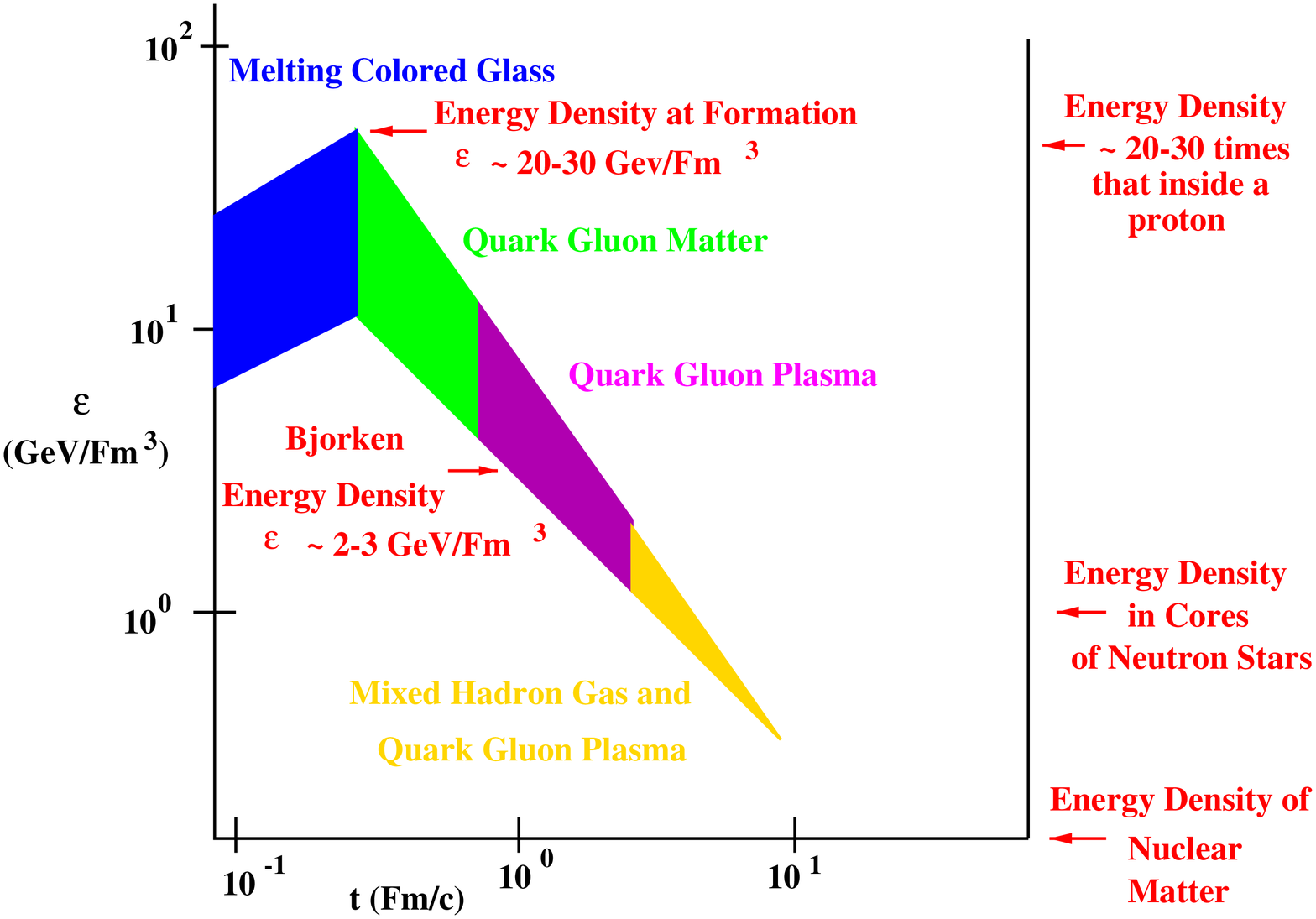}
        \caption{The evolution of a heavy ion collision at RHIC energy.
 }
\label{times}
    \end{center}
\end{figure}

The collision of two ultrarelativistic heavy ions can be visualized 
as the scattering of two sheets of colored glass, as shown in 
Fig. \ref{glasscoll}. \cite{mkw}-\cite{nkv}
At very early times after the collision the matter is at very high energy
density and in the form of a CGC.  As time goes on, the matter expands.
As it expands the density of gluons decreases, and gluons begin
to propagate with little interaction.
At later times, the interaction
strength increases and there is sufficient time for the matter to
thermalize and form a Quark Gluon Plasma.  This scenario is 
shown in Fig. \ref{times}, with realistic estimates for energy density
and time scales appropriate for the RHIC heavy ion accelerator.

\subsection{The Multiplicity}

\begin{figure}[ht]
    \begin{center}
        \includegraphics[width=0.50\textwidth,angle=270.]
{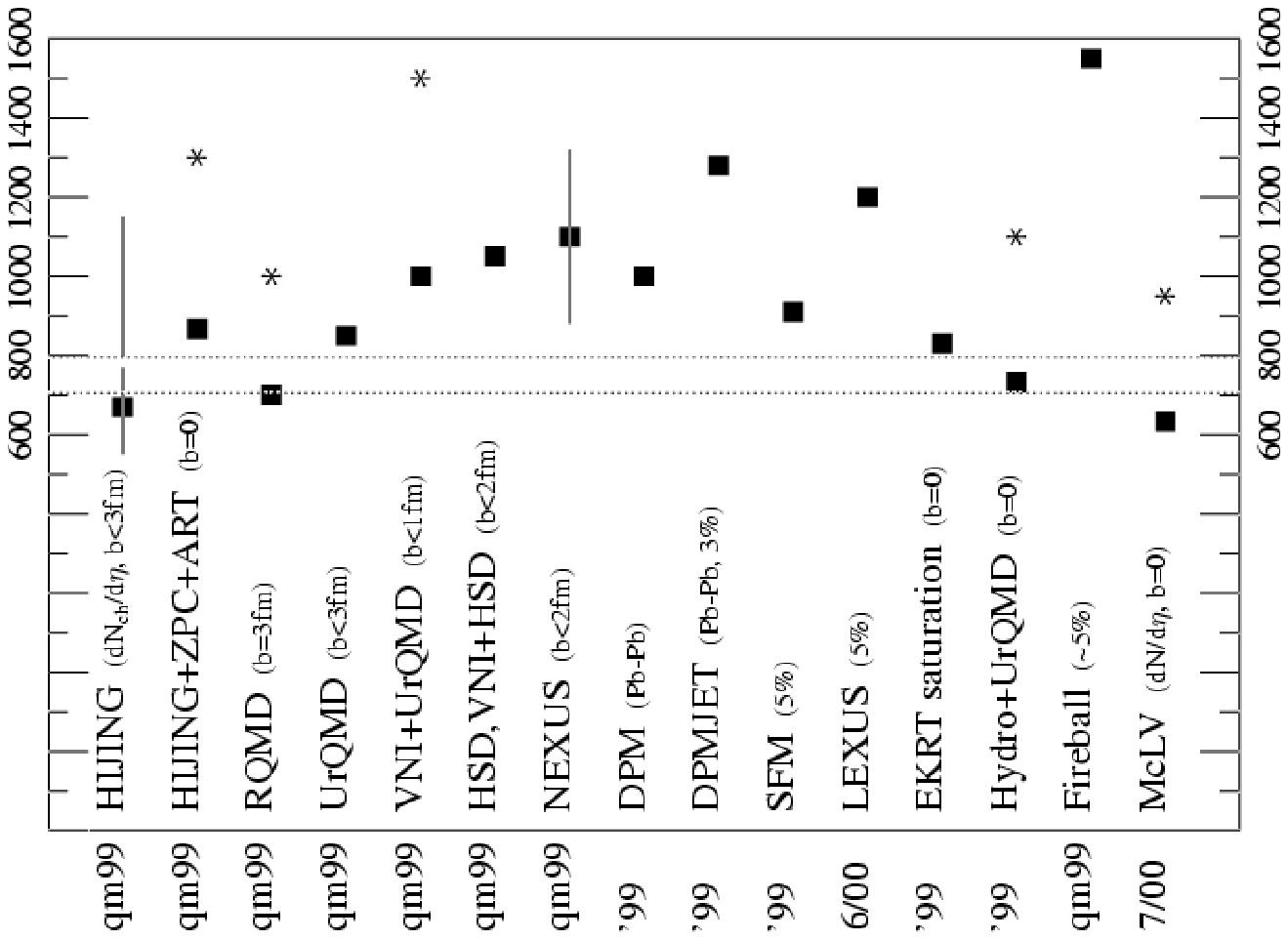}
        \caption{Predictions for the total multiplicity as measured at RHIC.
 The band is the experimentally measured region.  The CGC prediction
is marked McLV.}
\label{dndypred}
    \end{center}
\end{figure}

The CGC allows for a direct computation of the particle multiplicity
in hadronic collisions.  If one naively tries to compute jet production,
the total multiplicity is infrared divergent.  This follows because of
the $1/p_T^4$ nature of the perturbative formula for gluon production
\be
	{1 \over {\pi R^2}}{{dN} \over {dyd^2p_T}} \sim {1 \over \alpha_S}~
{{Q_{sat}^4} \over p_T^4}
\ee
In the CGC, when $p_T \le Q_{sat}$, this formula is cutoff.  This means that
the total gluon multiplicity goes as
\be
	{1 \over {\pi R^2}} {{dN} \over {dy}} \sim {1 \over \alpha_S}~ 
Q_{sat}^2
\ee
One can compute the proportionality constant and before the RHIC data appeared,
predictions were made for the gluon multiplicity.  In Fig. \ref{dndypred},
the predictions for the first RHIC run are presented.
The CGC was one of the few models which got the multiplicity correct.

\begin{figure}[ht]
    \begin{center}
        \includegraphics[width=0.80\textwidth]{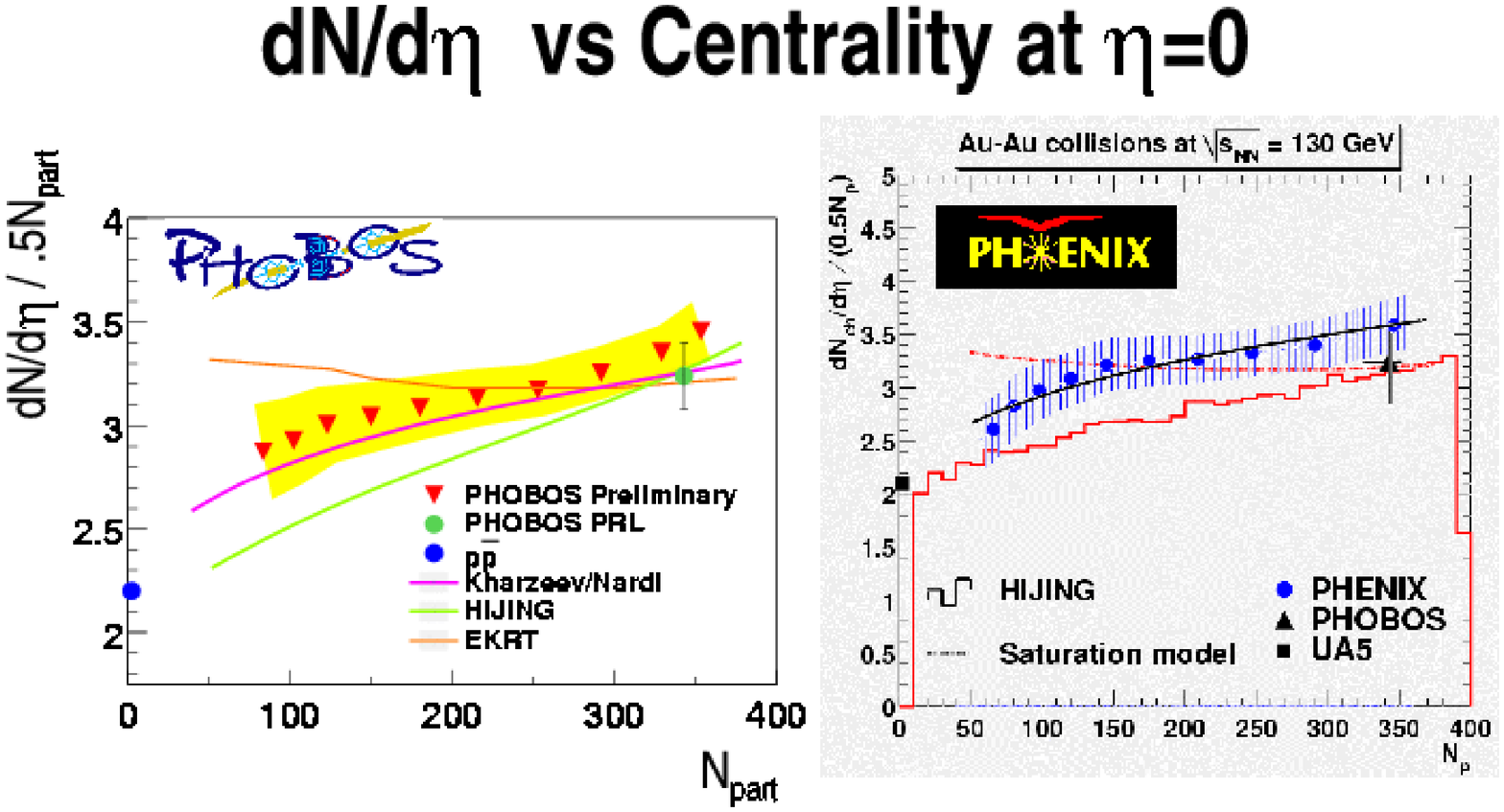}
        \caption{The total multiplicity as a function of the number
of participants as measured by Phobos and Phenix. 
 }
\label{glass1}
    \end{center}
\end{figure}

Also, the dependence of the multiplicity on the number of participants
can also be computed, realizing that the saturation momentum should be 
(for not too small x) proportional $N_{part}^{1/3}$.  This leads to
\be
	{{dN} \over{dy}} \sim {1 \over \alpha_S}
\ee
so that we have a very slow logarithmic dependence on the number of 
participants.  This was a prediction of the CGC and it agreed with experiment,
as shown in Fig. \ref{glass1}.\cite{ekrt}-\cite{phenixnpart}

One can go even further and compute the dependence of the multiplicity
on rapidity and centrality, and the transverse momentum distribution of 
produced hadrons by using CGC initial conditions matched together with
a hydrodynamic calculation which evolves the matter thought the
Quark Gluon Plasma.\cite{hirano}  The results describe the data 
remarkably.\cite{phobosdndy}-\cite{phenixdndpt}
\begin{figure}[htb]
    \centering
       \mbox{{\epsfig{figure=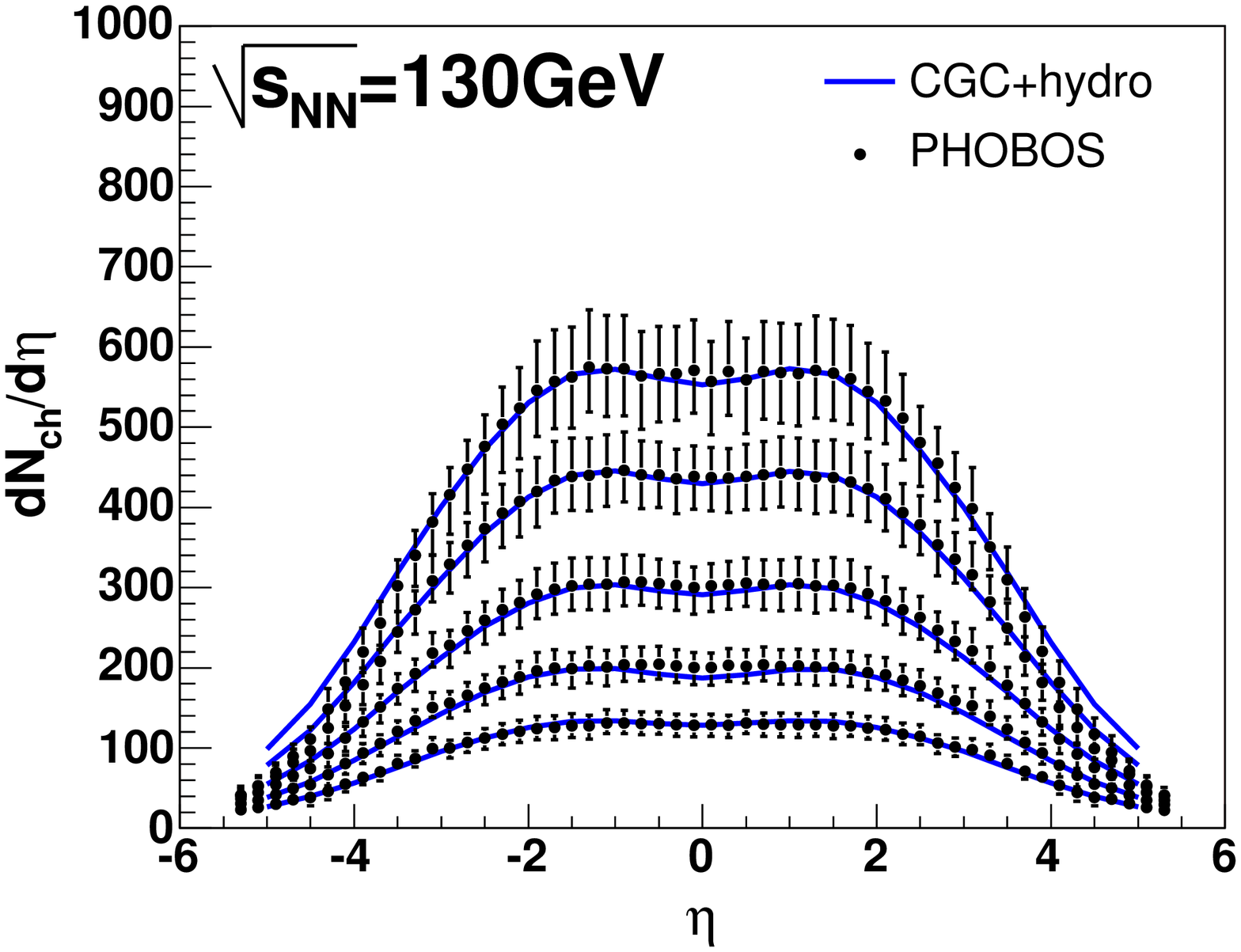,
        width=0.50\textwidth}}\quad
             {\epsfig{figure=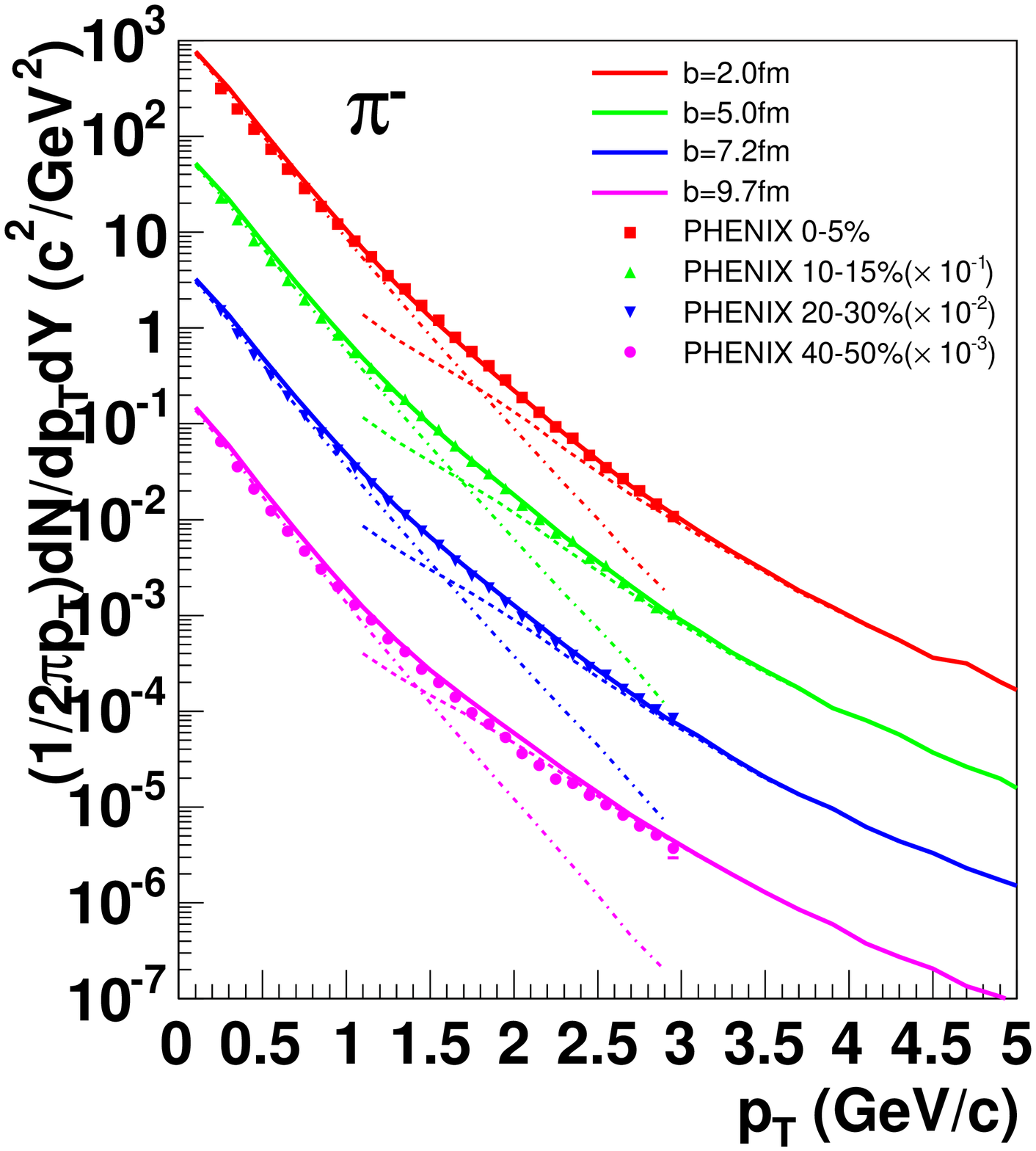,
        width=0.50\textwidth}}}
        \caption{Results of hydrodynamic simulation with
a CGC initial condition for the multiplicity as a function of y and
centrality, and for the transverse momentum distribution at zero
rapidity as a function of transverse momentum and centrality.}
        \label{hirano}
\end{figure}
This hydrodynamic simulation of Nara and Hirano does remarkably well.

\subsection{High $p_T$ Particles}

The early results from RHIC on gold-gold collisions revealed that the high
$p_T$ production cross sections were almost an order of magnitude below 
that expected for jet production arising from incoherent parton-parton 
scattering.\cite{expjetsuppression} This could be either due to initial state
shadowing of the gluon distribution inside the nuclei,\cite{klm}
or to final state jet 
quenching.\cite{thejetsuppression}  For centrally produced jets, the x of the
parton which produces a 5-10 GeV particle is of order $10^{-1}$, and this is
outside the region where on the basis of the HERA data one expects the
effects of the CGC to be important.  Nevertheless, nuclei might be different
than protons, so it is not a priori impossible.

The crucial test of these two different mechanisms is the comparison
of dA scattering to pp.  If there is suppression of jet in dA collisions, 
then it is an initial state effect.  The experiments were performed, and all 
there is little initial state effect for centrally produced 
jets.\cite{jetunsupp}
The suppression of centrally produced jets in AA collisions at RHIC is indeed
due to final state interactions, that is jet quenching.

This is not in contradiction with the existence of a CGC.  The particles
which control the multiplicity distribution in the central region are
 relatively
soft, and arise from $x \sim 10^{-2}$.  To probe such small $x$
degrees of freedom at high transverse momentum at RHIC requires 
that one go to
the forward region.\cite{dumitru}-\cite{gelis}

\begin{figure}[ht]
    \begin{center}
        \includegraphics[width=0.50\textwidth]{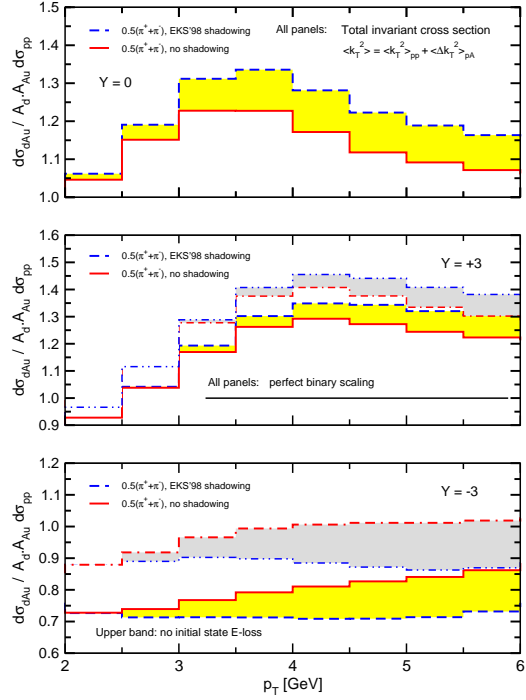}
        \caption{The expectations of classical multiple
scattering for the $p_T$ distribution in dA collisions.
 }
\label{vit}
    \end{center}
\end{figure}

If one uses naive Glauber theory to compute the effects of shadowing
by multiple scattering,
one expects that if one goes into the forward region of the deuteron, the probe
propagates through more matter in the nucleus.  This is because we
probe all of the gluons with $x$ greater than the minimum $x$ of the nucleus
which can be seen by the deuteron.  Going more forward makes this minimum 
x smaller.  Now multiple scattering will produce more particles 
at some intermediate value of $p_T$.  (At very high $p_T$, the
effects of multiple scattering will dissappear.)
This is the source of
the Cronin peak and it is expected to occur at $p_T$ of $2-4~GeV$.  Clearly
the height of this peak should increase as one goes more forward on the side
of the deuteron, and should increase with the centrality of the 
collision.\cite{vitev}
A result of such a computation is shown in Fig. \ref{vit}.

\begin{figure}[ht]
    \begin{center}
        \includegraphics[width=0.50\textwidth]{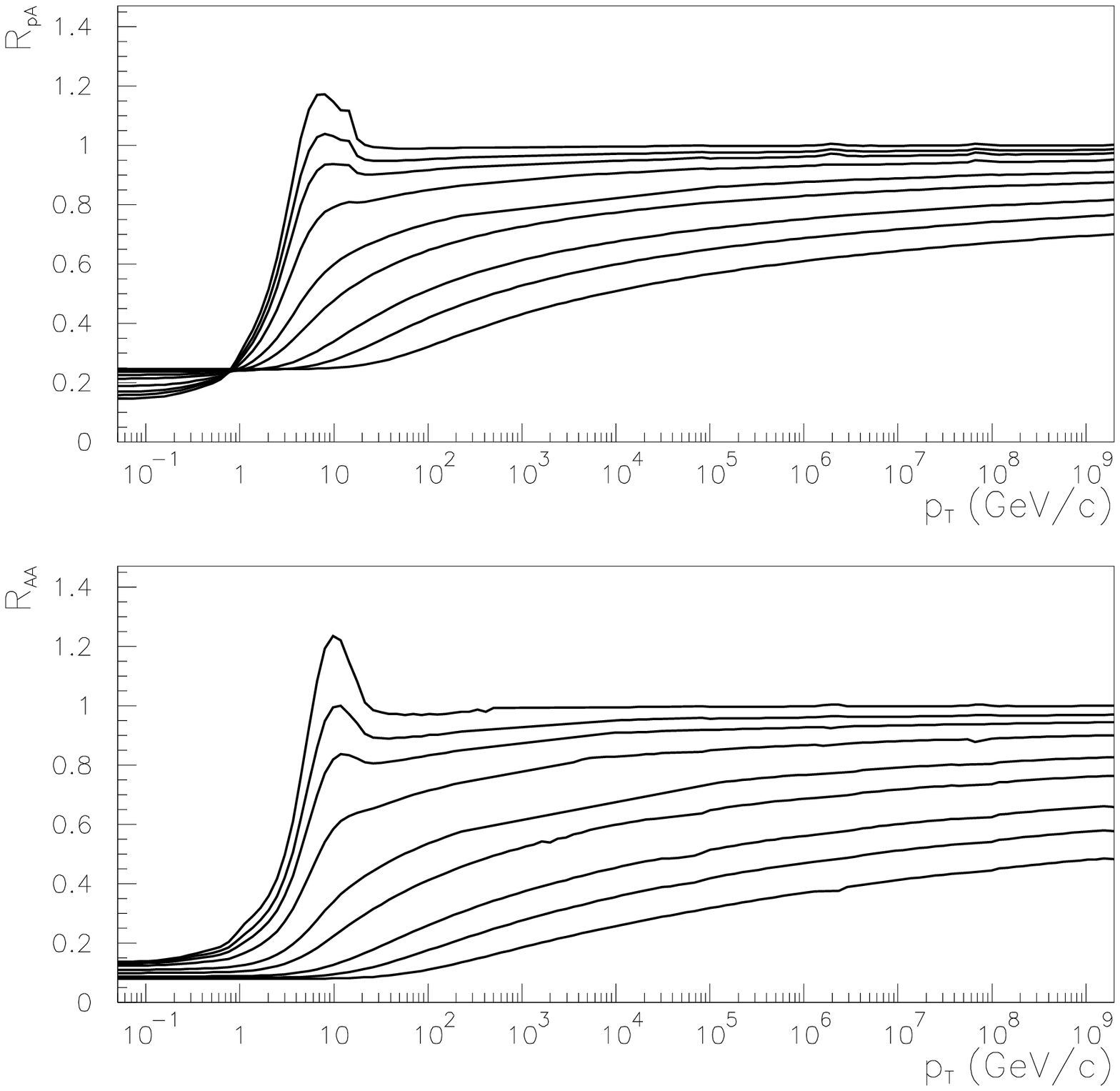}
        \caption{The gluon intrinsic gluon distribution function as a
function of $p_T$ for different pseudo-rapidities. 
 }
\label{kweid}
    \end{center}
\end{figure}

Classical rescattering effects are included in the computation of the 
properties
of the CGC.  There is another effect however and that is quantum evolution
generated by the renormalization group equations.  It was a surprise
that when one computed the evolution of the gluon distribution function 
including
both effects, the quantum evolution dominated.  This means that the height
of the Cronin peak, and the overall magnitude of the gluon distribution
decreased  as one went from backwards to forward angles.\cite{kkt}-
\cite{kw2}  The results of one such computation are shown in Fig. \ref{kweid}
It was also a surprise how rapid the effect set in.

\begin{figure}[htb]
    \centering
       \mbox{{\epsfig{figure=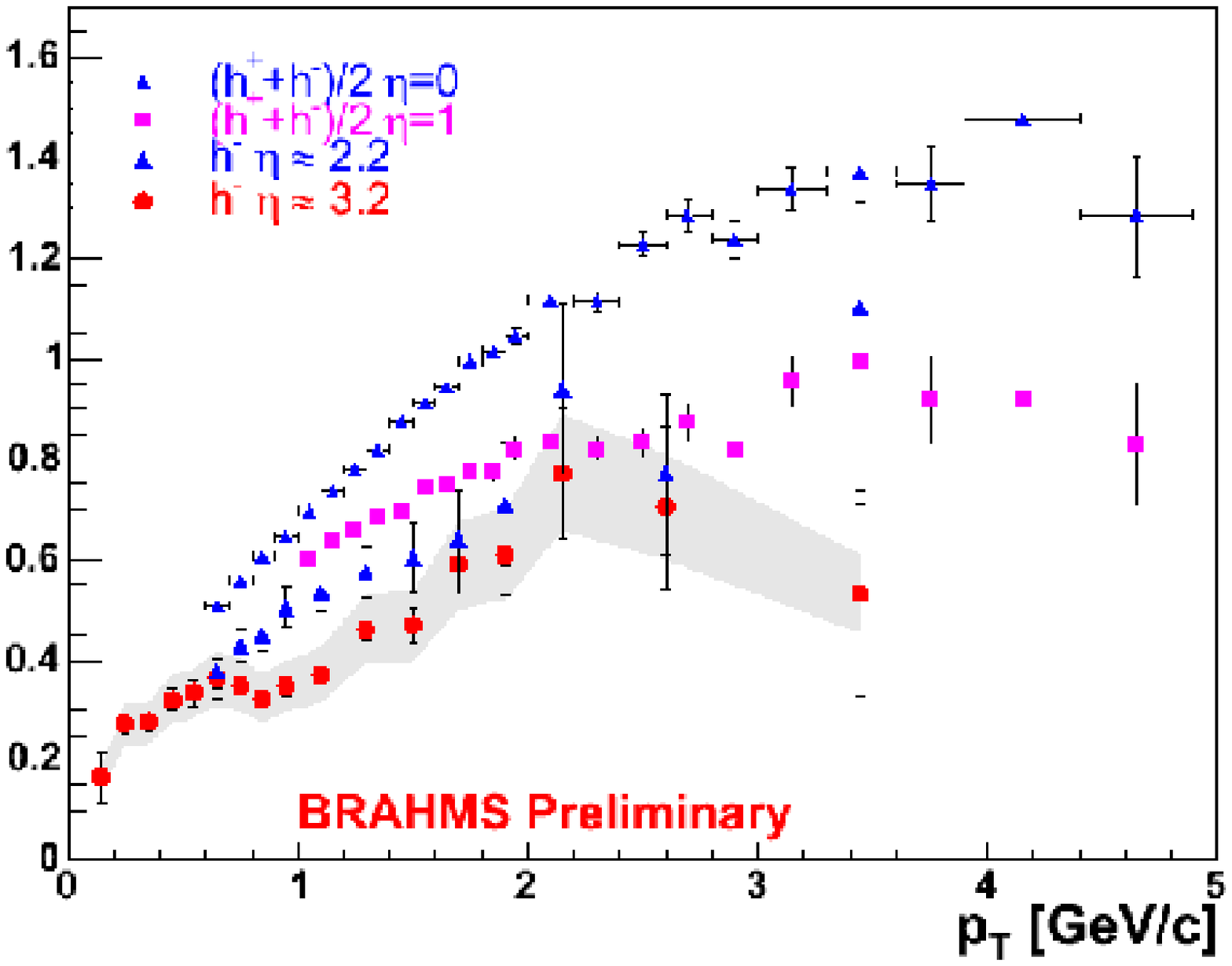,
        width=0.35\textwidth,angle=270.}}\quad
             {\epsfig{figure=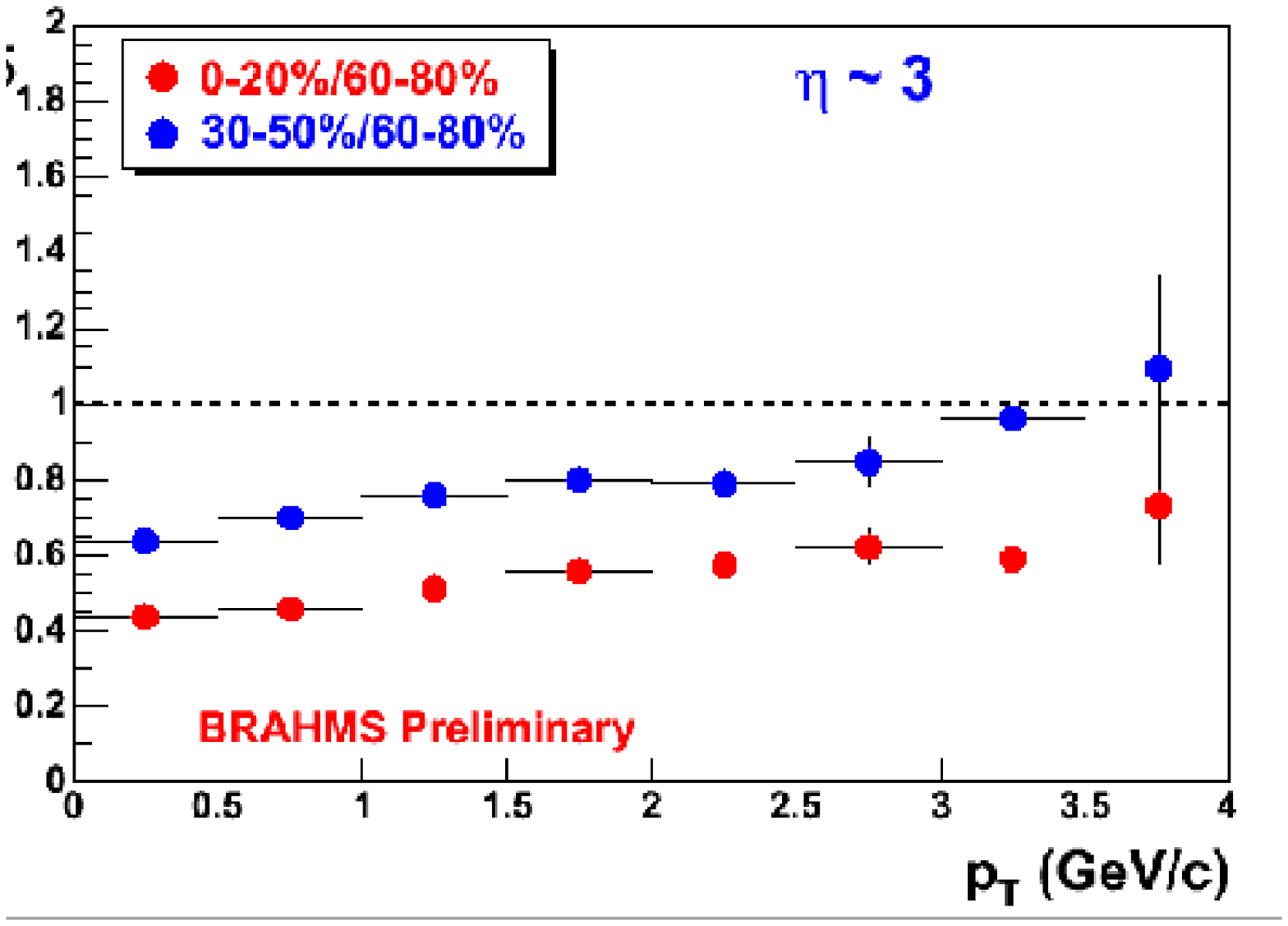,
        width=0.35\textwidth, angle=270.}}}
        \caption{(a) The  ratio $R_{CP}$ as a function of $p_T$
for various forward pseudorapidities. (b) The ratio $R_{CP}$ at a fixed 
forward pseudorapidity as a function of $p_T$ for less central and more
central dA collisions}
        \label{brahms}
\end{figure}

The Brahms experiment at RHIC recently presented data on the ratio
of central to peripheral transverse momentum 
distributions.\cite{brahmsda} The ratio
$R_{CP}$ is defined in such a way that if the processes were due to
incoherent production of jets, then $R_{CP} = 1$.  A value less than one
indicates suppression, and a value larger than one indicates a Cronin 
type enhancement.  The results for a variety  of forward angles for $R_{CP}$
as a function of $p_T$ is shown in Fig. \ref{brahms} a.  There is clearly
a decrease in $R_{CP}$ as one goes to forward angles, in 
distinction from the predictions of classical multiple scattering.  
The effect is very rapid in rapidity, as was expected from computations 
of the CGC.
In Fig. \ref{brahms} b, the ratio $R_{CP}$ is shown as a function of
$p_T$ for the forward pseudorapidity  $\eta \sim 3$ for less central and
more central events.  The ratio decreases for more central collisions,
against the expectation of classical multiple scattering and consistent with
the CGC hypothesis.

\begin{figure}[htb]
    
       {\epsfig{figure=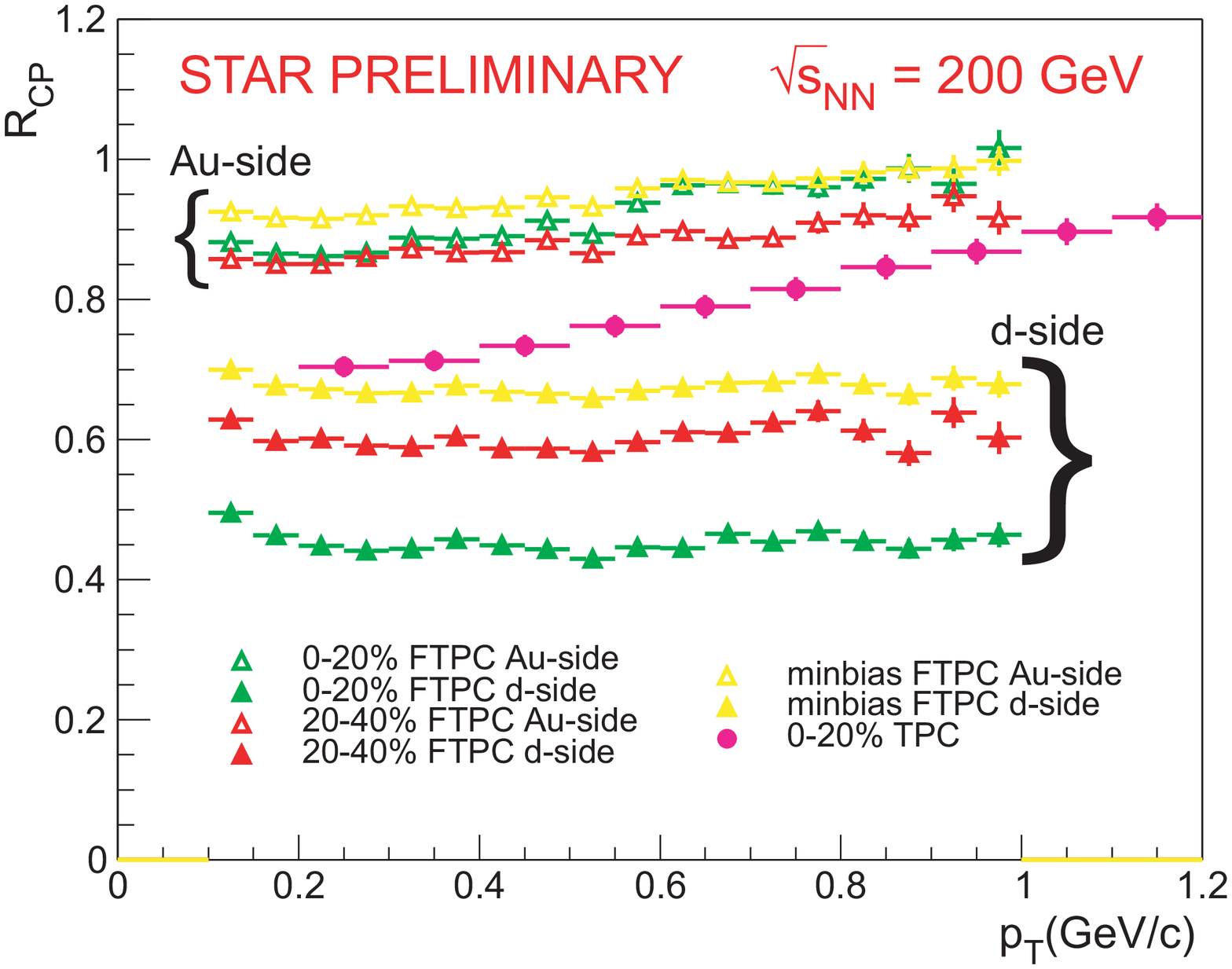,
        width=0.35\textwidth}}\quad
       { \epsfig{figure=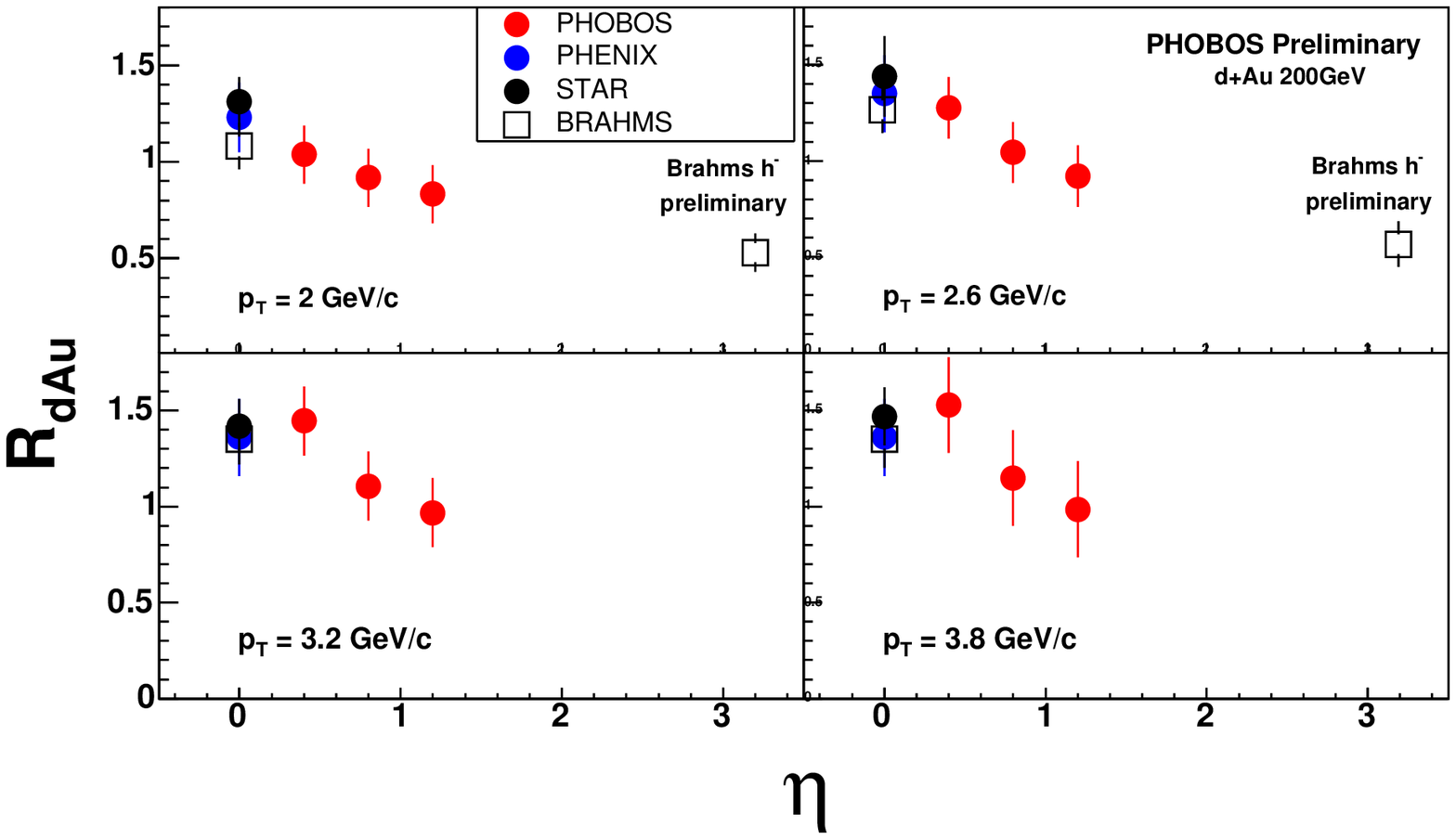,
        width=0.50\textwidth}}
        \caption{(a) The ratio $R_{CP}$ as a function of $p_T$ 
as measured in Star
for dA collisions at various pseudorapidities. (b) The ratio
$R_{dAu}$ as a function of pseudorapidity at fixed $p_T$ values
and different centralities }
        \label{stphda}
\end{figure}

\begin{figure}[htb]
    \centering
       \mbox{{\epsfig{figure=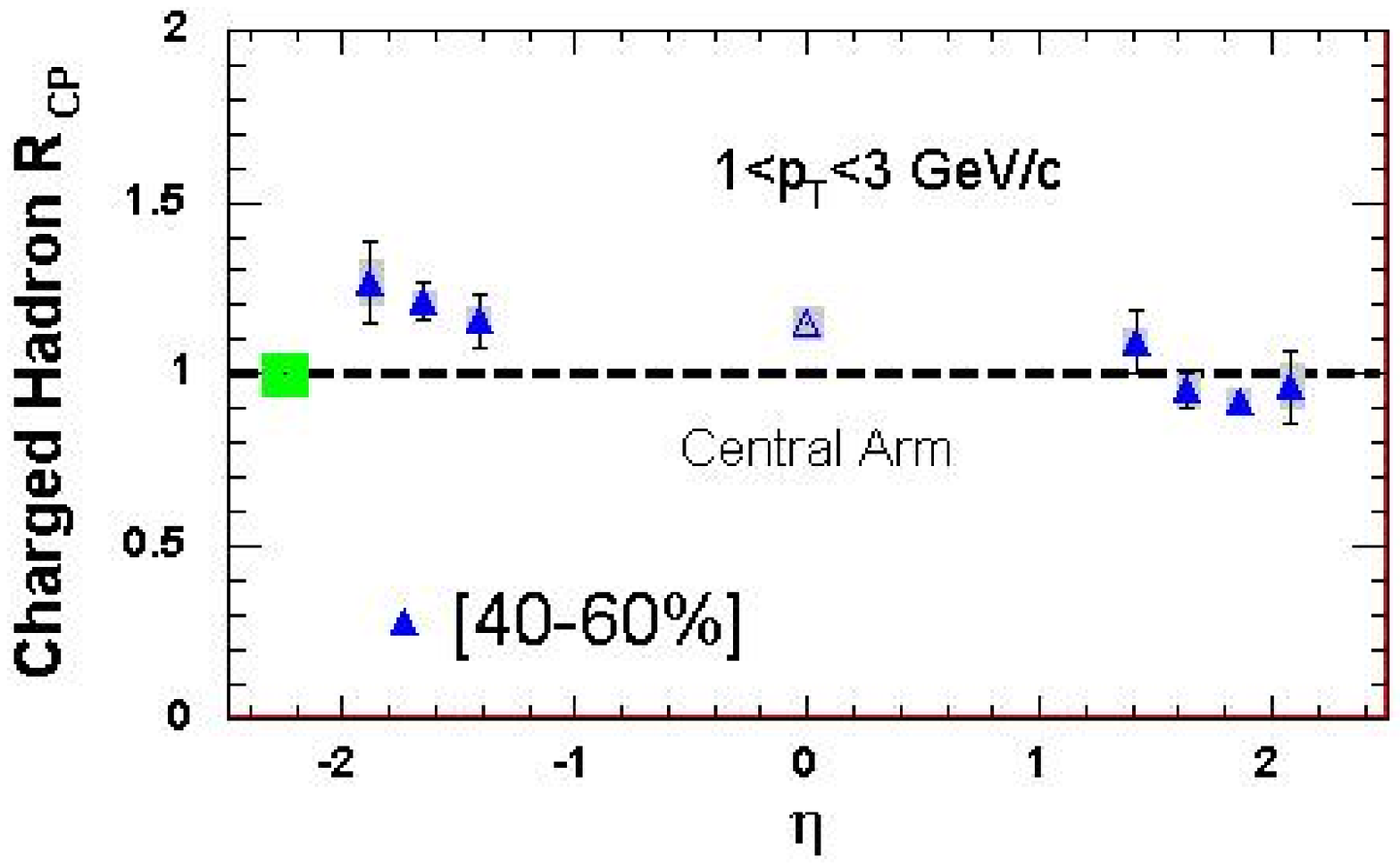,
        width=0.35\textwidth}}\quad
             {\epsfig{figure=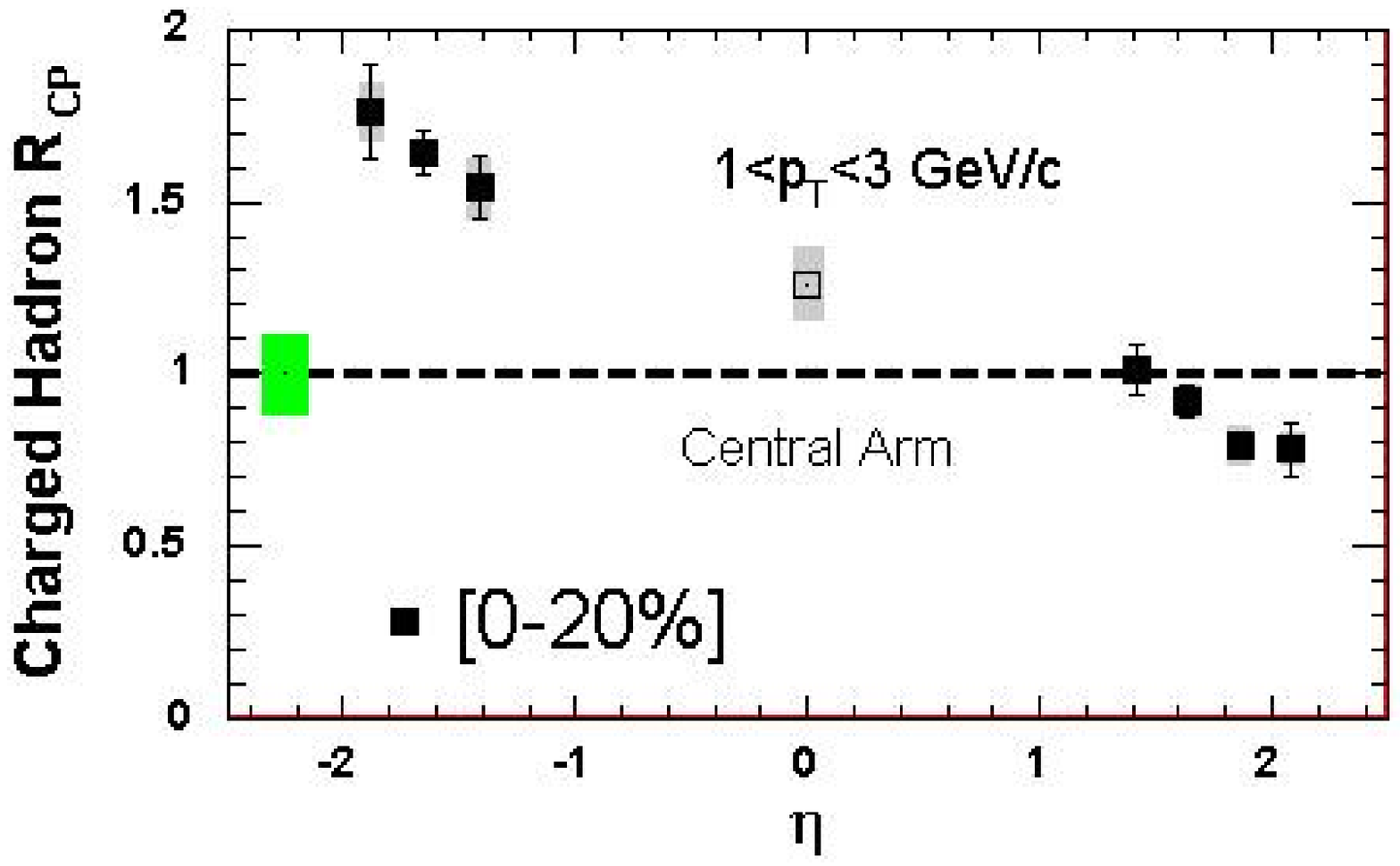,
        width=0.35\textwidth}}}
        \caption{The pseudorapidity distribution of the ratio $R_{CP}$
for stopped particles with $1~GeV \le p_T \le 3~GeV$  
at different centralities as measured in Phenix.}
        \label{phda}
\end{figure}

Similar results have been seen in the Star and Phobos 
experiment as shown in Fig.\ref{stphda} \cite{starda}-\cite{phobosda}
Phenix has also shown very dramatically the dependence upon
centrality of $R_{dAu}$ and pseudorapidity for hadron with
$1~GeV \le p_T \le 3~GeV$,\cite{phenixda} as seen in Fig. \ref{phda}  
This data beautifully illustrates the Cronin enhancement on the gold side
and the depletion on the deuteron side, and as well the dependence on 
centrality.  It appears that classical multiple scattering dominates
on the gold side, and quantum evolution on the deuteron side.  By a happy
coincidence, these effects nearly cancel in the mid-rapidity region, making
RHIC  a very good machine for studying QGP effects at midrapidity for
hard probes.  The $J/\Psi$ has also been measured in the Phenix experiment
and shows similar dependence on centrality and pseudorapidity as does
the hadron spectra\cite{phenixda}.

\subsection{Alternative Explanations}

Although classical multiple scattering cannot explain the effects seen
at RHIC in the forward $dA$ experiments, is it possible that some other
theory of shadowing can do it?  It has not been the expectation from a
number of computations which build in shadowing.\cite{miklos}  When 
any of these models are corrected to include the effects of classical multiple
scattering, they fail to describe the $dA$ data.\cite{accardi}-\cite{levai}
Nevertheless, someone will surely 
find some parameterization of shadowing based on somewhat 
shadowy assumptions and modeling which when combined with
classical multiple scattering will describe the data. 
The question which must be asked is how robust is the underlying model
which predicts these distributions.  The CGC provides a robust
theoretical framework based on first principles from QCD 
in which such shadowing and a variety of other effects are predicted.
The CGC also explains a variety of other phenomena, as we have seen.

One of the test of the CGC hypothesis will be the forward backward
correlations for jets in $dA$ collisions, where one of the jets
is at forward pseudorapidity.  If there is a CGC present, the
$p_T$ of the jet will be broadened on a scale of order of the 
saturation momentum.  At present, this measurement has not been 
reported\cite{bland}, nor have there been reliable theoretical computations.

\section{Summary}

My colleague Dima Kharzeev was quoted in the press as saying
about the CGC and the recent RHIC results
\begin{itemize}
\item{This is nothing short of a major discovery.}
\item{it's going to trigger a real revolution in nuclear physics}
\end{itemize}
By now, it should be clear why I support his position.
The measurements at RHIC which support the idea that there is ultradense
matter, and that at early times this matter has the properties predicted
for a Color Glass Condensate is without doubt a major discovery.  I think 
however in order to make the case for the Color Glass Condensate
compelling, one needs to supplement the RHIC data with electron scattering
data from HERA.  Additional tests will come from hard processes measured in 
LHC and potentially definitive tests from experiments at eRHIC. 

A revolution involves at least the following:
\begin{itemize}
\item{Revolutions involve major realignments of traditional relationships
between large groups of people.} 
\item{Friends try to kill one another.  Sometimes successfully.}
\item{There are bad consequences if you try to make a revolution and fail.}
\end{itemize}  

The kind of debate over the  CGC hypothesis is less whether
the CGC can describe phenomena observed in HERA and RHIC, but whether
or not there are alternative explanations.  These alternative explanations
often involve model computations, and do not try to unify the 
wide range of phenomena described in the talk.  It is always
very difficult to falsify a model because there is always some arbitrariness
in any model.  Also, models are a bit like a hydra:  If you rule out one 
model two new ones will appear.
On the other hand a theory, such as that of the CGC, 
must pass much more stringent
tests, and if it is wrong, it must be discarded.  Historically,
such debates are never resolved on the basis of whether a theory or a model
with free parameters can best describe data.  The conclusion will almost always
be that some model can.  
The reason that theories become accepted is because they
have a simple and unifying intellectual framework, that the arguments
which motivate the theory are compelling, and that it describes, within the
accuracy of various approximations, a wide variety of diverse
phenomena.

\section{Acknowledgements}
I gratefully acknowledge conversations with 
Edmond Iancu, Kazu Itakura, Miklos Gyulassy, Dima Kharzeev 
Genya Levin, and Raju
Venugopalan on the subject of this talk

This manuscript has been authorized under Contract No. DE-AC02-98H10886 with
the U. S. Department of Energy.

\end{document}